\documentclass[9pt,journal]{IEEEtran}
\usepackage{graphicx} 
\usepackage[hidelinks]{hyperref}      
\usepackage{cite}            
\usepackage{mathrsfs}        
\usepackage{amsfonts}        
\usepackage{amsmath}
\usepackage{amssymb}
\usepackage{subfigure}
\usepackage{multirow}
\usepackage{algorithm}
\usepackage{algorithmic}
\usepackage{color}
\usepackage{amsmath}
\usepackage{array}
\usepackage{algorithm}
\usepackage{algorithmic}
\usepackage{stfloats}

\newcommand{\RNum}[1]{\uppercase\expandafter{\romannumeral #1\relax}}

\begin{document}

\title{A Demand-Supply Cooperative Responding Strategy in Power System with High Renewable Energy Penetration}

\author{Yuanzheng Li, {\it Senior Member, IEEE},
		 	Xinxin Long, 
		 	Yang Li, {\it Senior Member, IEEE},
		 	Yizhou Ding, {\it Student Member, IEEE},
		 	Tao Yang, {\it Senior Member, IEEE},
		 	Zhigang Zeng, {\it Fellow, IEEE}

\thanks{This work was supported by the Key Scientific and Technological Research
Project of State Grid Corporation of China under Grant 5108-202315041A-1-1-ZN.}
     
\thanks{Y. Z. Li, X. X. Long and Z. G. Zeng are with School of Artificial Intelligence and Automation, Huazhong University of Science and Technology, Wuhan 430074, China (Email: Yuanzheng\_Li@hust.edu.cn, xinxin\_l@hust.edu.cn, zgzeng@hust.edu.cn).}

\thanks{Y. Li is with School of Electrical Engineering, Northeast Electric Power University, Jilin 132000, China (Email: liyang@neepu.edu.cn).}

\thanks{Y. Z. Ding is with School of China-EU Institute for Clean and Renewable Energy, Huazhong University of Science and Technology, Wuhan 430074, China (Email: jasondean910@outlook.com).}

\thanks{T. Yang is with State Key Laboratory of Synthetical Automation for Process Industries, Northeastern University, Shenyang 110819, China (Email: yangtao@mail.neu.edu.cn).}

}

	\maketitle
	\begin{abstract}
	
	Industrial demand response (IDR) plays an important role in promoting the utilization of renewable energy (RE) in power systems. However, it will lead to power adjustments on the supply side, which is also a non-negligible factor in affecting RE utilization. To comprehensively analyze this impact while enhancing RE utilization, this paper proposes a power demand-supply cooperative response (PDSCR) strategy based on both day-ahead and intraday time scales. The day-ahead PDSCR determines a long-term scheme for responding to the predictable trends in RE supply. {However, this long-term scheme may not be suitable when uncertain RE fluctuations occur on an intraday basis. Regarding intraday PDSCR, we formulate a profit-driven cooperation approach to address the issue of RE fluctuations.} In this context, unreasonable profit distributions on the demand-supply side, would lead to the conflict of interests and diminish the effectiveness of cooperative responses. {To mitigate this issue, we develop multi-individual profit distribution marginal solutions (MIPDMSs) based on satisfactory profit distributions, which can also maximize cooperative profits. Case studies are conducted on a modified IEEE 24-bus system and an actual power system in China.} The results verify the effectiveness of the proposed strategy for enhancing RE utilization, via optimizing the coordination of IDR flexibility with generation resources.

	\end{abstract}


\IEEEpeerreviewmaketitle

\section{Introduction}
\IEEEPARstart{C}{urrently}, human development is significantly constrained by energy and environmental crises. Renewable energy (RE) is considered as a solution for replacing conventional energy sources, due to its environmental friendliness \cite{ref1,ref2,ref3}. {For instance, Wind power (WP) is a significant form of RE, which has developed rapidly in recent years, especially in China \cite{ref4}. However, the uncertainty of RE, such as the unpredictable and significant fluctuations,} may adversely affect the secure and cost-effective operation of the power system. {Therefore, there is a need to address this challenge regarding RE uncertainty \cite{ref5,ref6}}.

Demand response (DR) is considered as an effective method to mitigate RE power fluctuations, and RE utilization could be enhanced. Within the DR program, electricity customers accord to long-term pricing schemes or financial incentives, to flexibly schedule their power demand. {Notably, industrial production enterprises (IPEs) constitute a significant portion of global power demand, accounting for more than 50$\%$ of the total demand worldwide \cite{ref7}.} Therefore, to enhance the management of IPE demand, Ref. \cite{ref8} investigates industrial demand response (IDR) for minimizing generation costs and RE curtailments. {Furthermore, Ref. \cite{ref10} analyzes the potential for implementing existing IDR programs among industrial consumers, in order to augment RE utilization.}

{The previous studies treat IPEs as independent dispatch entities, which overlooks their interactions with the utility center (e.g., electricity manager) within the IDR program. To analyze these interactions, Ref. \cite{ref11} examines the impact of dynamic electricity pricing on demand response, involving 802 businesses across 34 commercial and industrial categories. On this basis, the interactions among multi-level market participants are modeled, according to {\emph {Stackelberg}} game theory \cite{ref12}.} These participants include the utility center, load aggregators, energy storage operators, etc. Furthermore, a programming-multi-verse distributed algorithm is introduced to optimize trading strategies among the utility center and load aggregators \cite{ref13}.

{In summary, the aforementioned researches mainly investigate centralized power dispatch within the day-ahead time scale. In this scenario, the utility center holds a priority in the process of electricity sales and price negotiations \cite{refI14}. This priority enables it to facilitate IDR more effectively via controlling power supply for transactions.} Moreover, unpredictable fluctuations in RE supply or power demand potentially impact power balance and stable power transmission on the intraday time scale. In addition to the established electricity contracts, the utility center generally offers additional financial incentives to encourage further shifts in IPE demand. This supplementary IDR exists outside of contractual limitations, allowing IPEs to autonomously determine their responses based on self-interests \cite{ref14}.

In contrast to centralized studies, decentralized dispatch relies on autonomous IDR to manage power fluctuations. {It means that all IDR participants act in a self-interested and profit-driven manner. More specifically, participants concentrate on maximizing the responding income, rather than strictly adhering to demands of the utility center. Therefore, the optimal response outcome depends on the effective cooperation among all participants.} Furthermore, an effective cooperation is dependent on a good financial incentive mechanism. Conversely, an inappropriate financial mechanism may result in excessive incentives for certain participants \cite{refre1}, which will weaken the motivation of other participants and even degrade response outcomes. On the other hand, the absence of penalties may encourage extreme selfish behavior and consequently adversely affect total interest \cite{refre2}. {This implies that a participant may prioritize their own interests over collective interests, leading to the conflict of interests among multiple individuals.}

In order to standardize the decentralized response of participants, Ref. \cite{ref15} establishes a punishment rule via the Cartel mechanism and repeated games. They are used to regulate the cooperation in the IDR program and avoid the conflict of interests. {From a profit perspective, Ref. \cite{ref16} formulates a non-cooperative game model among DR aggregators. Based on incomplete information, this study determines the DR share of each aggregator by maximizing the revenue of utility center.} Ref. \cite{ref17} considers the cooperation model as a more suitable approach for describing this decentralized interest relationship. Then, a cooperative IDR scheme is proposed to achieve power demand management and the distribution of cooperative profits.

On the other hand, {employing real-time thermal power response is a conventional approach for mitigating power fluctuations \cite{refre4}. However, due to the increasing replacement of thermal power with RE, this strategy has become increasingly challenging \cite{refre5}. The reason is this unilateral response relies on the adjustable thermal power on a large scale. In this context, the collaborative response of both thermal power and industrial load presents a valuable idea. The basis is that collaborative adjustments on both the supply and demand sides can effectively balance power shortage or excess, making it a more adaptable approach for handling RE fluctuations \cite{rereref1,rereref2}.} In power management within electricity markets of UK and US, both thermal power adjustments and demand response mechanisms are utilized to maintain power balance \cite{refre3}. {However, these processes are primarily practiced as simple unilateral transactions, for suppliers or power consumers.}

Therefore, from the above literature review, we note the following research gaps. 
({\emph {i}}) Concerning intraday response and financial mechanism, the research scope of the conflict of interest is limited to the demand side in the decentralized response. {However, it is important to clarify that the demand shifts will correspondingly cause supply adjustments. Therefore, a rational response strategy should not be limited to the allocation of resources on the demand side alone. It should also encompass power dispatch on the supply side, and coordinate the interests of multiple individuals on the demand-supply side.} Our work focuses on studying the comprehensive management of conventional power supply and industrial demand, and establishing a more organic cooperation strategy. ({\emph {ii}}) For the demand-supply interest, it is unilateral to focus solely on the influence of financial commendation on the decentralized response. {This perspective overlooks the potential cost increase due to the industry load response. In this context, participants may not necessarily obtain benefits, compared with the original scheme. Therefore, it is necessary to comprehensively take into account both cost variations and financial incentives, in order to accurately quantify the final profit.}

In our work, the coupling relation of IDR and thermal power dispatch is firstly established. Then, the related multi-individual interests on the demand-supply side are studied based on day-ahead and intraday response strategies. In the day-ahead time scale, the utility center coordinates the long-term scheme of thermal unit commitment and electricity price. This is to promote the demand-supply responses to the varying trend of RE supply. Specifically, we establish a bi-level multi-objective model to characterize this centralized cooperation. In the intraday time scale, an incentive scheme promotes decentralized cooperative response on the demand-supply side, for addressing the RE fluctuation. In this context, we quantify final profits by the cost difference between the intraday and day-ahead schemes, to characterize the demand-supply interest. As mentioned previously, { improper distributions of cooperative profits would lead to the conflict of interests among multiple parties, potentially diminishing the effectiveness of cooperative responses. Consequently, we study the fair and rational distribution within cooperative responses, to address this conflict. To be concluded, contributions of this paper are summarized as follows.}

\begin{itemize}
  \item[(1)] The power demand-supply cooperative responding (PDSCR) strategy is proposed to promote RE utilization, which primarily adopts the schemes in different dispatch periods (i.e., day-ahead and intraday), and the corresponding coordination of demand-supply side interests.
  
  \item[(2)] {A bi-level multi-objective optimization (BMO) is formulated to determine the day-ahead PDSCR scheme. In addressing the predictable variations in RE supply trends, the BMO considers decision priorities, demand-supply costs, and transmission risk within the power system operation.}
  
  \item[(3)] A profit-driven multi-individual cooperative model is developed in intraday PDSCR for addressing the RE fluctuation. Feasibility of the cooperative model is verified through corresponding proof. On this basis, considering the fair and rational distribution, the multi-individual profit distribution marginal solutions (MIPDMSs) are deduced to deal with the conflict of interests caused by improper distributions.
  
\end{itemize}




The rest of this paper is organized as follows. Section II introduces the overview of the PDSCR strategy. Section III establishes a bi-level multi-objective optimization model regarding the day-ahead PDSCR. Section IV formulates a cooperation model for the intraday PDSCR. Section V focuses on the case study and simulation results. Finally, conclusions are drawn in Section VI. {Also, Table I presents the correspondence between the acronym and the original word.}

\vspace{-0.3cm}
\begin{table}[h]\scriptsize
   \renewcommand{\arraystretch}{1.5}
	\begin{center}
		\caption{Acronym Table}
		\begin{tabular}{m{0.8cm}<{\centering}m{2.4cm}<{\centering}|m{0.8cm}<{\centering}m{3.0cm}<{\centering}} \hline\hline	
			Acronym &  Full word  & Acronym & Full word  \\   
			\hline
			IDR   & Industrial demand response     & MPEC        & Mathematical programming with equilibrium constraints   \\   
			PDSCR   & Power demand-supply cooperative responding    & KKT      & Karush-Kuhn-Tucker   \\          
		    RE   & Renewable energy    & NBS   & Nash bargaining solution \\ 
		    WP   & Wind power    & MIOMM   & Multi-individual optimal marginal model   \\ 
		    IPE   & Industrial production enterprises    & MIPDMS   & Multi-individual profit distribution marginal solution   \\ 
		    PMP   & Parallel manufacturing process    & PF   & Pareto front   \\ 
		    ASC   & A security coefficient    & BMO   & Bi-level multi-objective optimization   \\ 
			\hline\hline	
		\end{tabular}
	\end{center}
	\vspace{-0.7cm}
\end{table}

\section{Overview Of Power Demand-Supply Cooperative Responding Strategy}

In this section, we first introduce the composition of the power system in Subsection A. Next, in Subsection B, we explain properties of two industrial demand responses. Then, Subsection C proposes the PDSCR strategy across different time scales and explains their corresponding strategy purposes and interest relationships.

\subsection{Composition of Power System}

\begin{figure}[htb]
	\vspace{-0.2cm}
	\centering
	\includegraphics[scale=0.33]{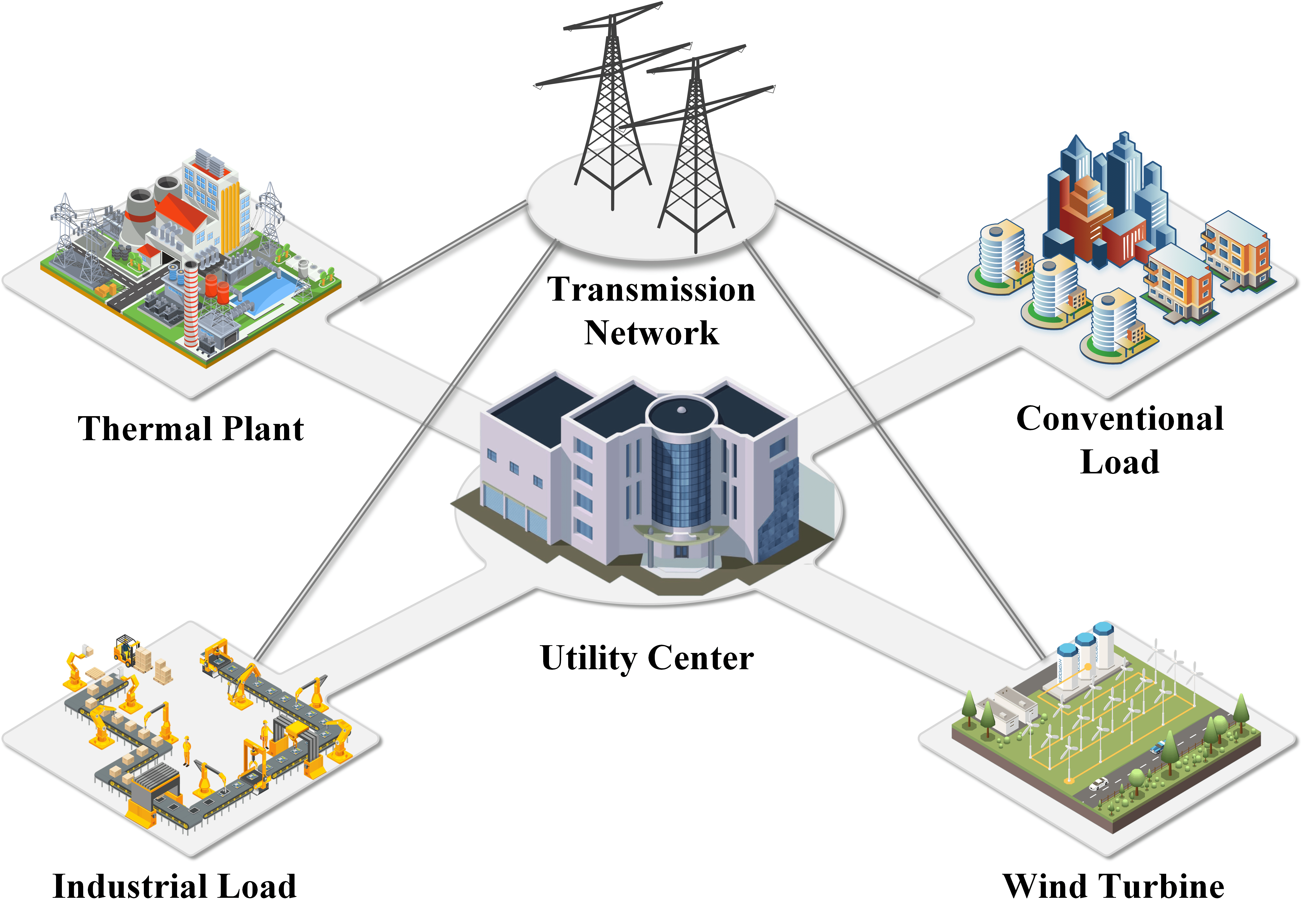}
	\vspace{0cm}
	\caption{A power system consisting of conventional thermal plants, transmission network, and industrial loads (i.e., IPE).}\label{Fig1}
	\vspace{-0.1cm}
\end{figure}

The specific power system considered in this paper is depicted in Fig. \ref{Fig1}. It comprises wind turbines, industrial loads, conventional thermal plants, and the power transmission network.

 {In light of energy and environmental considerations, there is a need to enhance wind power utilization on the supply side. However, it will potentially introduce overload risks in the transmission network \cite{rereref5, rereref6}. To address this concern, it is imperative to comprehensively promote the flexibility of industrial power demand and corresponding thermal power dispatch.}

\begin{figure*}[htb]
	\vspace{-0.2cm}
	\centering
	\includegraphics[scale=0.37]{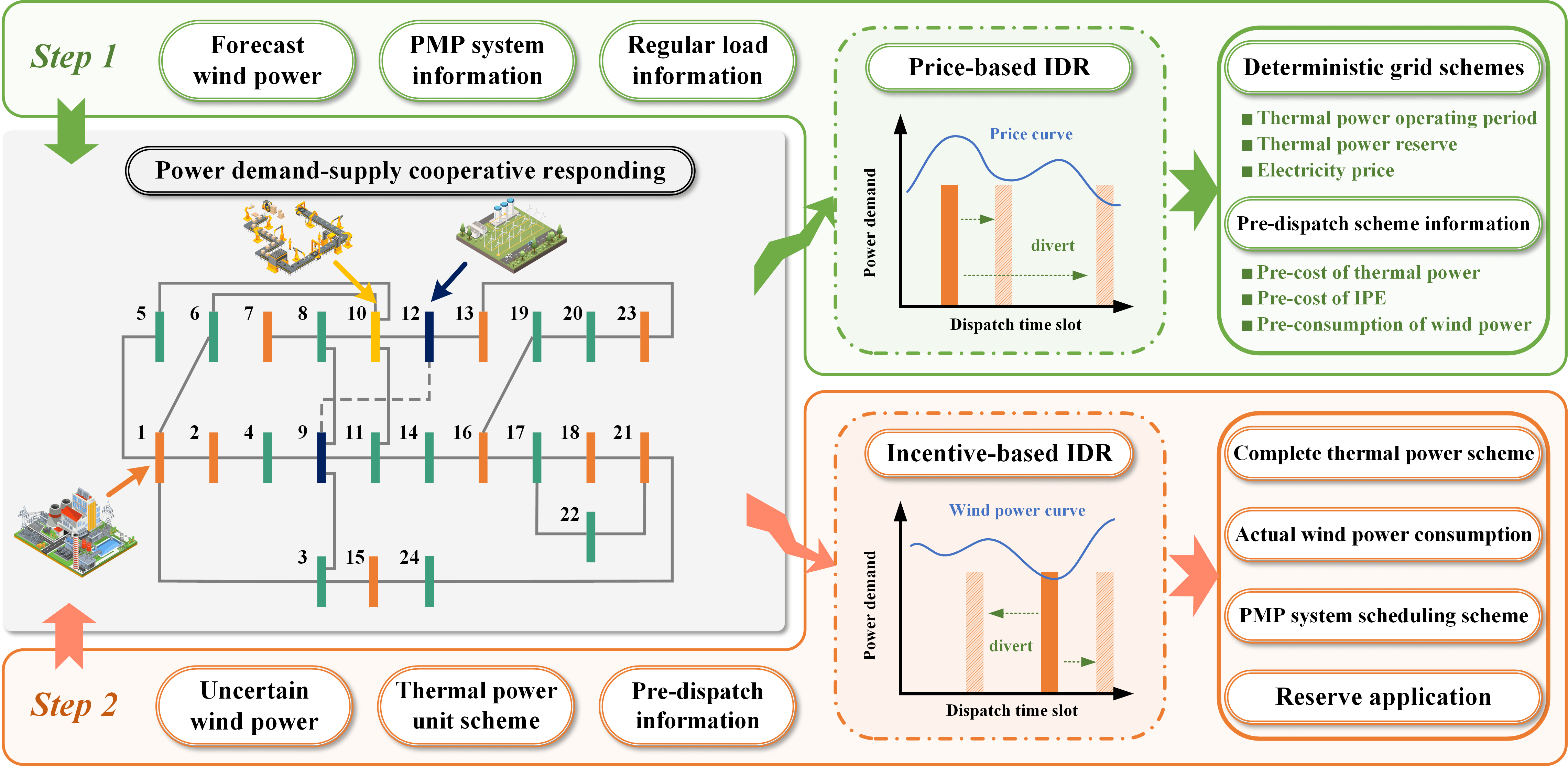}
	\vspace{0.1cm}
	\caption{Procedure of the proposed PDSCR strategy.}\label{Fig4}
	\vspace{-0.4cm}
\end{figure*}

\subsection{Two IDR programs}

As shown in Fig. \ref{Fig3}, we incorporate both price-based and incentive-based IDR programs into PDSCR and analyze components of IPE electricity fares. The price-based IDR program relies on the contracted electricity prices. Specifically, IPEs can reduce production cost by shifting their loads to low-price periods and avoiding power usage during high-price periods, as illustrated in Fig. \ref{Fig3}(a). Consequently, the price variations among different time periods encourage industrial demand shifting, that is, absorbing excess wind power or preventing power shortages. {This program serves as a foundational approach for basically determining intraday electricity demand. However, unforeseen fluctuations in wind power supply would lead to power imbalances. In practice, unilateral price adjustments within the contract are not accepted by IPEs on the intraday basis, because such adjustments may disrupt production and incur additional cost.

\begin{figure}[htb]
	\vspace{-0.2cm}
	\centering
	\includegraphics[scale=0.45]{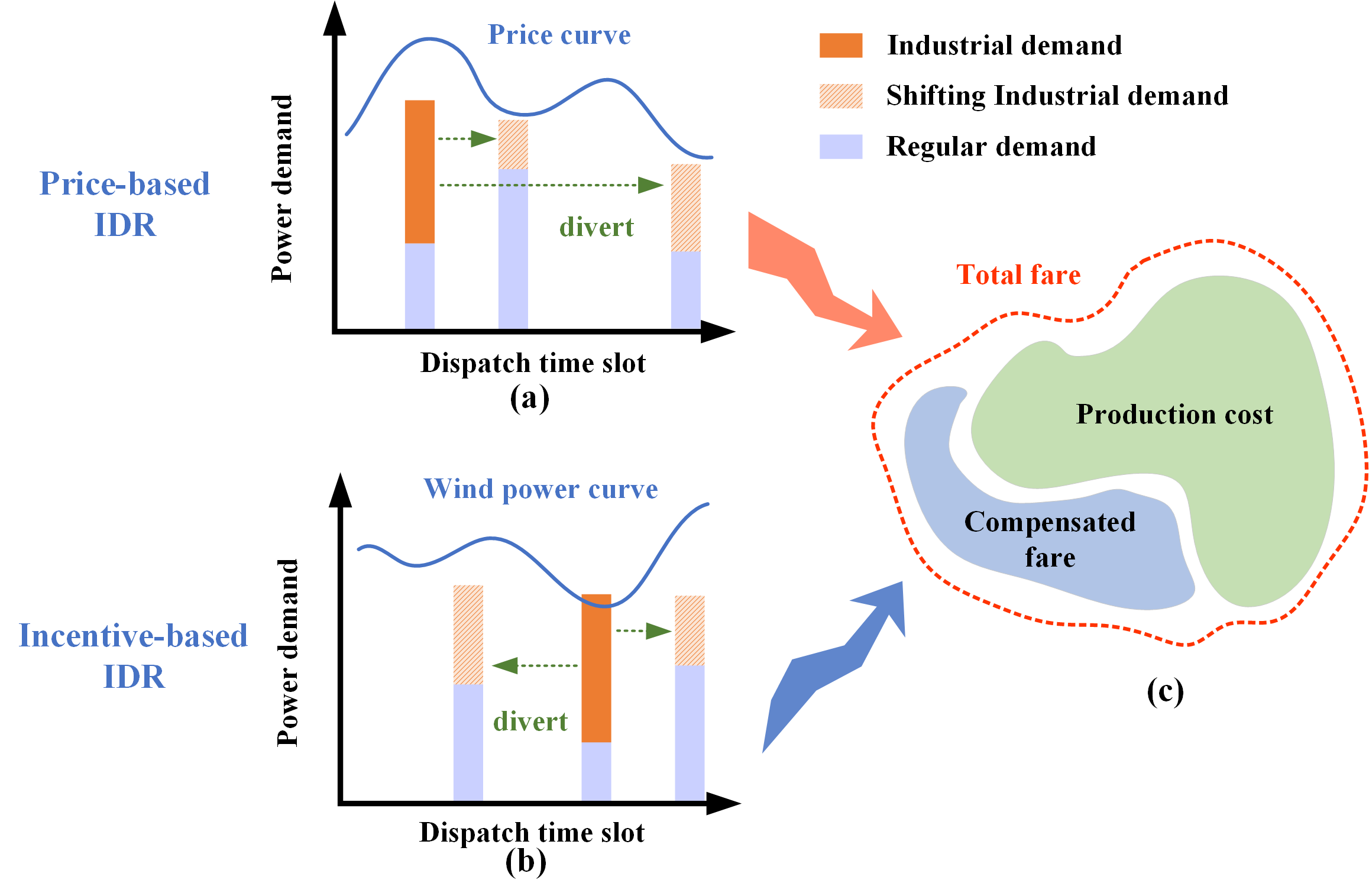}
	\vspace{0cm}
	\caption{Two types of IDRs, i.e., price-based and incentive-based IDRs.}\label{Fig3}
	\vspace{-0.2cm}
\end{figure}

On the other hand, driven by additional financial incentives, industrial customers further adjust their demand to address intraday power imbalances, known as incentive-based IDR. Specifically, financial compensations or penalties drive industrial power demand to increase during peak periods or decrease during valley periods of wind power supply, as depicted in Fig. \ref{Fig3}(b). For electricity transactions, the predictable supply and demand on the market need to be cleared ahead of the day. For uncertain events, the real-time response outside of the contract serves as a supplement. Based on this frame, the IPE production costs can be calculated as the difference between the total fare and the compensation fare, as shown in Fig. \ref{Fig3}(c).}

\subsection{Day-ahead and Intraday PDSCR Strategies}

{In Fig. \ref{Fig4}, the PDSCR strategy is divided into two steps, corresponding to the day-ahead and intraday time scales, respectively. In {\emph {Step}} 1, the utility center holds decision-making authority over electricity pricing and the scheme for thermal unit operations in day-ahead PDSCR. This can be viewed as a centralized pre-dispatch strategy. {In this process, the utility center firstly formulates the price-based IDR program and the day-ahead thermal unit commitment. Subsequently, the IPE gives a production plan to determine industrial power demand.} The relationship among these entities is described based on a {\emph {Stackelberg}} game framework, formulated as a bi-level multi-objective optimization. Moreover, the day-ahead scheme encompasses thermal generators' ON/OFF states, reserves, electricity prices, etc.

{However, issues like branch overloads or power shortages may arise if actual wind power significantly deviates from its predictions. To address these challenges, {\emph {Step}} 2 focuses on enhancing the response to accommodate power deviations, i.e., the actual thermal power output and IDR program would be updated. In terms of the utility center's scheme, electricity pricing and thermal unit operation schemes remain unchanged. On this basis, the intraday PDSCR strategy emphasizes autonomous cooperation and offers corresponding financial incentives, including the thermal power compensation and incentive-based IDR.} Specifically, the intraday PDSCR strategy determines thermal power, wind power integration, industrial production behavior, and reserve usage.

\section{Day-ahead PDSCR strategy via Bi-level Multi-objective Optimization Model}

In this section, we introduce the day-ahead PDSCR strategy, which concentrates on the interest relationship of centralized power dispatch. At first, the industrial production process is formulated in Subsection A. It is a flexible optimization and responding model. Then, based on the {\emph {Stackelberg}} game, we establish the day-ahead PDSCR strategy as the bi-level multi-objective optimization model.

\subsection{Modeling of PMP Production Optimization Dispatch}

As flexible industrial loads, the parallel manufacturing process (PMP) system is introduced in this subsection. Within the PMP, IPE can achieve the optimal price cost by optimizing the PMP schedule. In this paper, the optimization model of the PMP system is established, and its process with ${\rm R}$ procedures is shown in Fig. \ref{Fig5}.

\begin{figure}[!h]
	\vspace{-0.2cm}
	\centering
	\includegraphics[scale=0.73]{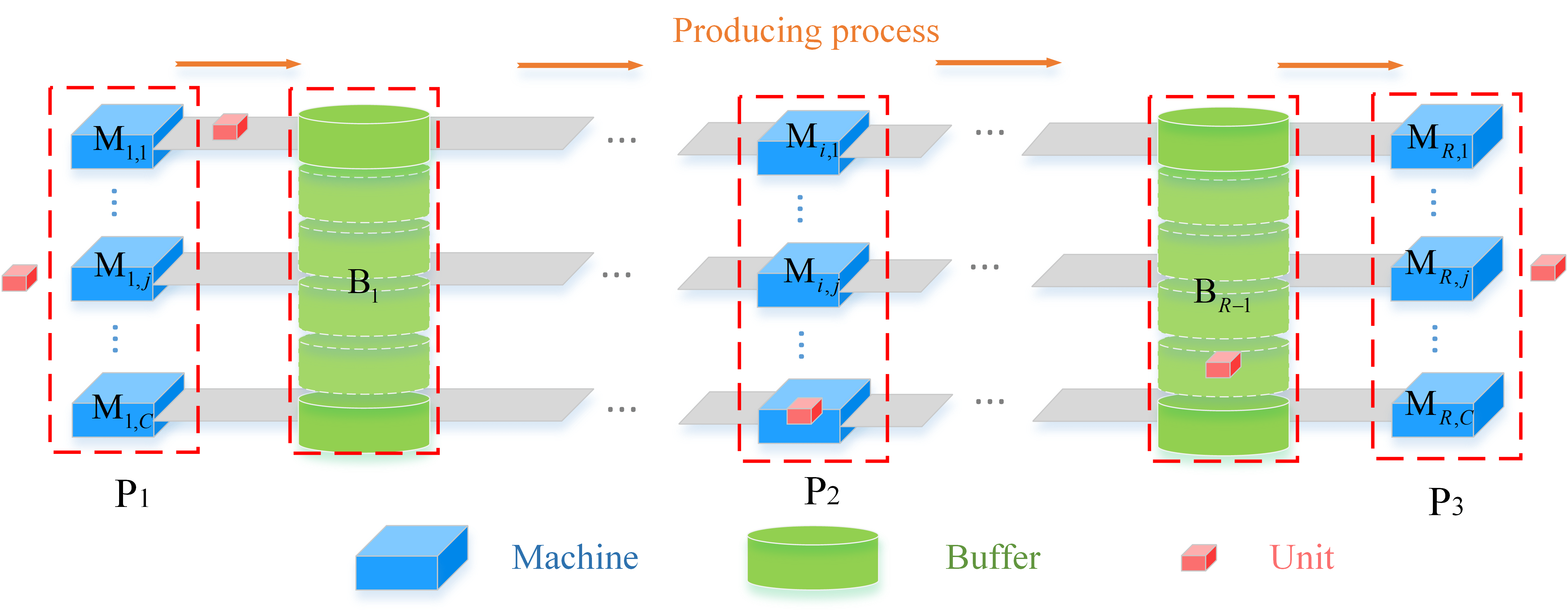}
	\vspace{-0.2cm}
	\caption{Production network structure of a PMP system.}\label{Fig5}
	\vspace{-0.2cm}
\end{figure}


The $i^{th}$ production procedure is denoted as ${\rm P}_i$ ($i=$ $1,2,...,{\rm R}$), in which there exist ${\rm C}$ machines, ${\rm M}_{i,1}, {\rm M}_{i,2},…, {\rm M}_{i,{\rm C}}$, and there are buffers (${\rm B}_1, {\rm B}_2,…, {\rm B}_{{\rm R}-1}$) between two adjacent procedures, which are used to store projects and provide flexibility for the PMP system to conduct adjusting operations. The process constraint of the PMP system is explained in (A.1)-(A.7) of Appendix A. Accordingly, in the production cycle consisting of ${\rm T}$ time intervals, the dispatch model of IPE is formulated as follows.
\begin{equation}\label{eq8}
\begin{array}{l}
\begin{aligned}
{\rm{min }} \quad &{J_{{\rm{PMP}}}} = \sum\limits_{t = 1}^{\rm{T}} {\left( {{F_{{\rm{P}},t}} + {F_{{\rm{B}},t}}} \right)}  + {{\rm{C}}_{{\rm{FIXED}}}}\\
{\rm{s.t.}} \quad &{F_{{\rm{P}},t}} = \sum\limits_{i = 1}^{\rm{R}} {{\alpha _t} \cdot {N_{{\rm{P}},t,i}}}  \cdot {c_{{\rm{P}},i}}\\
&{F_{{\rm{B}},t}} = \sum\limits_{i = 1}^{{\rm{R}} - 1} {{\alpha _t} \cdot {N_{{\rm{B}},t,i}}}  \cdot {c_{{\rm{B}},i}} \\
&\rm{(A.1)-(A.7)}
\end{aligned}
\end{array}
\end{equation}
where ${N_{{\rm{P}},t,i}}$ is the quantity of processed projects of procedure ${{\rm{P}}_i}$ and time slot ${t}$. Similarly, the storage quantity of buffer ${{\rm{B}}_i}$ is denoted as ${N_{{\rm{B}},t,i}}$. ${c_{P,i}}$ and ${c_{{\rm{B}},i}}$ are cost coefficients of ${P_i}$ and ${B_i}$, ${\alpha _t}$ is the electricity price in $t$ slot, and ${{\rm{C}}_{{\rm{FIXED}}}}$ is the fixed cost. The optimal electricity cost is consider as the interest on the demand side.

\vspace{-0.2cm}
\subsection{Formulations of Bi-level Multi-objective Optimization}

In day-ahead PDSCR, the utility center coordinates thermal power supply and industrial demand to respond to the peak-valley trend of wind power supply. On this basis, it is also necessary to guarantee cost optimality on the demand-supply side. However, this cooperative response is not a simple multi-objective optimization problem, and is with a bi-level optimization existing in the order of priority among multiple individuals.

\begin{figure}[htb]
	\vspace{-0.3cm}
	\centering
	\includegraphics[scale=0.48]{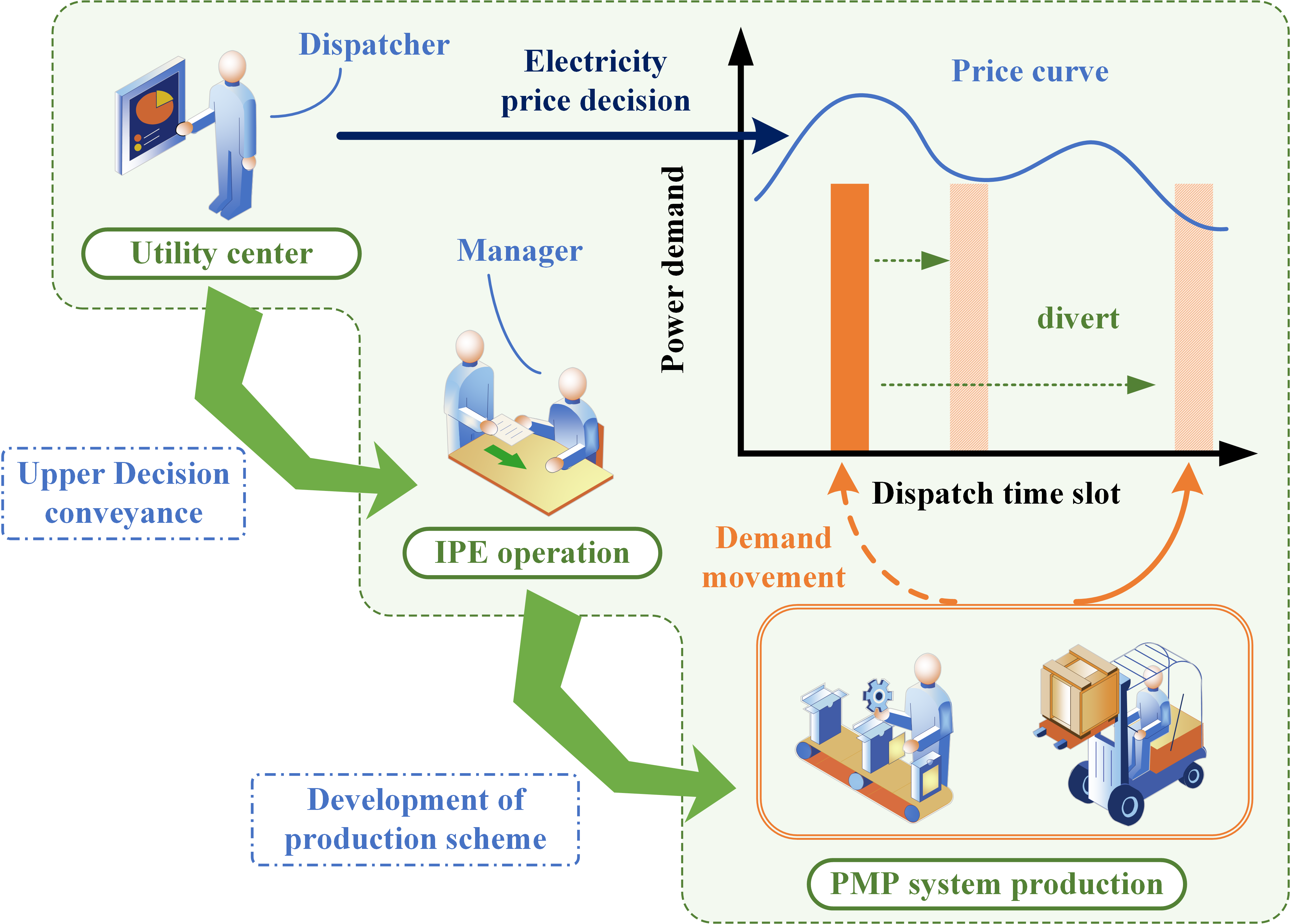}
	\vspace{0cm}
	\caption{Workflow of squential desicion and coordination.}\label{Fig7}
	\vspace{-0.5cm}
\end{figure}

Specifically, on the supply side, the utility center devotes to minimizing outputs of operational thermal power units and enhancing wind power utilization. This amounts to optimizing the overall cost associated with thermal power generations. On the demand side, the utility center employs price-based IDR to impact industrial load shifting. {As shown in Fig. \ref{Fig7}, the utility center establishes the long-term electricity pricing scheme based on the potential for IDR adoption. Then,  industrial customers determine the optimal production schedule considering electricity prices and their specific industrial requirements. This process represents a sequential decision-making approach. The {\emph{Stackelberg}} theory introduces the concept of priority among game participants. This priority dictates the relative importance of each participant's objectives. Therefore, this game model is proved effective in characterizing the hierarchical game relationships \cite{refre8}.} On this basis, we assume a two-individual game distinguished between leader 1 and follower 2, and their utility function and behavior are ($F_1,b_1$) and ($F_2,b_2$). That is, equilibrium $b_1^*$ in {\emph {Starkelberg}} game is as (\ref{eqin1}).
\begin{equation}\label{eqin1}
F_1^* = \mathop {\max }\limits_{{b_2} \in {O_2}(b_1^*)} {F_1}(b_1^*,{b_2}) = \mathop {\min }\limits_{{b_1} \in {B_1}} \mathop {\max }\limits_{{b_2} \in {O_2}({b_1})} {F_1}({b_1},{b_2})
\end{equation}
where $B_1$ represents the leader behavior set. $O$ stands for the following response function for given $b_1^*$. If $O$ is the optimal response function ${b_2^*} \in {O_2}(b_1^*)$, the two behavior $({b_1^*},{b_2^*})$ will be {\emph {Starkelberg}} equilibrium. { Just as in the mathematical programming with equilibrium constraints (MPEC), the optimal response function can be transformed as constraints, according to the convex nature of follower optimization. This transformation is achieved by deducing it through the {\emph {Karush-Kuhn-Tucker}} (KKT) condition \cite{refre6,ref19}. Therefore, relevant constraints in the bi-level optimization model are established as (\ref{eqI9}). The specific constraints is formulated as (A.8)-(A.14) in Appendix A.}
\begin{equation}\label{eqI9}
{\nabla _y}{L_{{\rm{PMP}}}} = 0,\quad \lambda  \circ {f_{neq}}(x,y) = 0
\end{equation}
where ${{\cal L}_{{\rm{PMP}}}}$ is the {\emph {Lagrange}} function, ${f_{neq}}(x,y)$ is inequality constraint vector, the corresponding dual variable vector $\lambda$ is non-negative, and $\circ$ is the {\emph {Hadamard}} product \cite{rereref3}. Then, the following is the proposed bi-level and multi-objectives optimization model in day-ahead PDSCR, where the decision variable of utility center is $x=[{s_{t,p}},{P_{{\rm{T}},t,p}},{R_{{\rm{T}},t,p}},{P_{{\rm{W}},t,l}},{{\alpha _t}}]$ and the IPE variable is $y=[{N_{{\rm{P}},t,i}},{N_{{\rm{B}},t,i}}]$. $x$ includes thermal On/Off status ${s_{t,p}}$, power outputs ${P_{{\rm{T}},t,p}}$, reserves ${R_{{\rm{T}},t,p}}$, wind power integration ${P_{{\rm{W}},t,l}}$, electricity price ${{\alpha _t}}$ for time $t$, thermal generator $p$, wind unit $l$, and number of thermal power units $N_{\rm G}$. $y$ includes numbers of processed projects ${N_{{\rm P},t,i}}$ and buffer ${N_{{\rm{B}},t,i}}$ for time $t$ and process $i$. In addition, a security coefficient (ASC) is introduced to quantify the risk based on the available transfer margin of the branch. This margin of a branch is described by the difference between the transmitted power and its rated value during the dispatch periods. In (\ref{eq16}), ${C_{{\rm{ASC}}}}$ is defined as the summation of minimum margins of all the branches and periods. It is noteworthy that the margin should be maximized.
\begin{equation}\label{eq16}
{C_{{\rm{ASC}}}} = \mathop {\min }\limits_j \left( {\frac{{T{P_{j,\max }} - \mathop {\max }\limits_t \left( {T{P_{t,j}}} \right)}}{{T{P_{j,\max }}}}} \right) \cdot 100\% 
\end{equation}
where $T{P_{t,j}}$ and $T{P_{j,\max }}$ denote transmitted power and the maximum margin of branch $j$ $(j=1,2,…, b)$, respectively. Generally, the more wind power utilization means the less thermal cost. On this basis, we formulate the thermal cost and the $C_{\rm{ASC}}$ as the upper optimization objectives, because the utility center is in dominance. Here, the bi-objectives optimization would obtain the optional optimal solutions, i.e., Pareto solutions. The reason is that different preferences for the bi-objectives will cause different optimal solutions. Pareto optimality means there is no feasible solutions $x_1$ and $x_2$ where objectives $J_i (x_1) \ge J_i (x_2), \forall i \in \{1,2\}$ with $J_i (x_1) > J_i (x_2), $ for some $i$, i.e., there is no complete advantage among all the Pareto solutions. Therefore, the decision-making involves the trade-off for Pareto solutions.

On the other hand, the industrial electricity cost is concerned by IPE. In this paper, this electricity cost is formulated as the lower optimization objectives, the performance of which depends on the upper decision. The specific bi-level optimization model is established as (\ref{eq17}), the supplemental constraints of the thermal power and grid are shown in (B.2)-(B.14) of Appendix B. {This is an MPEC model, where $\min \; [{J_1},{J_2}]$ represents the optimal leader's utility, and $\arg \min \; {J_{{\rm{PMP}}}}$ represents the optimal follower's utility. The optimal response function is defined as the constraint (\ref{eqI9}), specific forms of which are provided in equations (A.8)-(A.14).}
\begin{equation}\label{eq17}
\begin{array}{l}
\begin{aligned}
\min \quad &[{J_1},{J_2}]\\
{\rm{s.t.}} \quad &{J_1} = \sum\limits_{t = 1}^{\rm{T}} {\sum\limits_{k = 1}^{{N_{\rm{G}}}} {[F({P_{{\rm{T}},t,p}}) + {S_{{\rm{T}},t,p}}]} } \\[1.1mm]
&{J_2} = 1 - {C_{\rm {ASC}}}\\[1.1mm]
&\arg \min \quad {J_{{\rm{PMP}}}}\\[1.1mm]
&{(1) - (4), \rm{(B.2)-(B.14)}}
\end{aligned}
\end{array}
\end{equation}
where $F({P_{{\rm T},t,i}})$ and $S_{{\rm T},t,p}$ represent thermal generation cost function and starting cost. In this multi-objective problem, the Pareto solutions can be obtained by $\varepsilon$-constraint optimization method \cite{ref20}. 

\section{Intraday PDSCR strategy in multi-individual cooperation model}

In this section, the profit-driven cooperation among multiple individuals is considered on the demand-supply side. Firstly, the model of intraday PDSCR involving financial commendation is established in Subsection A. To study the distribution of cooperative profits, Subsection B proposes the multi-individual profit distribution marginal solution.

\subsection{Modeling of Intraday PDSCR}

The intraday PDSCR strategy is based on the combination of thermal power dispatch and incentive-based IDR, and is autonomous and profit-driven. Considering the thermal power commendation and the incentive strategy within the IDR scheme, the updated electricity cost function of the PMP system could be reconstructed as (\ref{eq20}), which introduces the incentive ${S_t^{\rm {P,R}}}$. The updated cost function of the thermal plants is formulated as (\ref{eq21}), where the electricity sales revenue $J_3$ and thermal power commendation ${S_{{\rm T},t,k}^{\rm R}}$ are introduced.
\begin{equation}\label{eq20}
{J_3} = \sum\limits_{t = 1}^{\rm T} {\left( {F_{{\rm{P}},t}^{{\rm{UP}}} + F_{{\rm{B}},t}^{{\rm{UP}}}} \right)}  - \sum\limits_{t = 1}^{\rm T} {S_t^{\rm {PMP,R}}}  + {{\rm{C}}_{{\rm{FIXED}}}}
\end{equation}
\begin{equation}\label{eq21}
\begin{array}{l}
\begin{aligned}
{J_4} = {\rm{ }}\sum\limits_{t = 1}^{\rm{T}} {\{ \sum\limits_{i = 1}^{{N_{\rm{G}}}} {[F(P_{{\rm{T}},t,p}^{{\rm{UP}}}) + {S_{{\rm{T}},t,p}}]} }  + C_{{\rm{outres}}}^t - \sum\limits_{k = 1}^{{N_{\rm{T}}}} {S_{{\rm{T}},t,k}^{\rm{R}}} \}  - {J_3}
\end{aligned}
\end{array}
\end{equation}
where $N_{\rm T}$ represents the number of thermal plants, and ${C_{\rm {outres}}^t}$ stands for the purchasing cost of reserve electricity. UP denotes the updated decision, where ${s_{t,p}}$, ${R_{{\rm{T}},t,p}}$ and ${{\alpha _t}}$ follow the day-ahead decision. The sum of ${S_{{\rm T},t,k}^{\rm R}}$ and $S_t^{\rm {P,R}}$ should be proportional to the increase of wind power utilization $\Delta {P_{\rm {W},t}}$. 
\begin{equation}\label{eq19}
{s_t} \cdot \Delta {P_{\rm {W},t}} = S_t^{\rm {P,R}} + \sum\limits_{i = 1}^{{N_{\rm{T}}}} {S_{{\rm T},t,k}^{\rm R}}
\end{equation}
where ${s_t}$ is the commendation constant. It means that multi-individual cooperation on the demand-supply side could produce more financial commendation. Evidently, for each independent period $t$, the total profit ${\psi _t}(x,y)$ is constructed by the thermal cutbacks, the commendation of wind power utilization, and the reserve power cost.
\begin{equation}\label{eq22}
\begin{array}{l}
{\psi _t}(x,y) = {[{J_{\rm {PMP}}} + {J_1} - {J_3} - {J_4}]_t}\\[1mm]
 = \sum\limits_{i = 1}^n {F({P_{{\rm P},t,p}})}  - \sum\limits_{i = 1}^n {F(P_{{\rm P},t,p}^{\rm{UP}})} + {s_t} \cdot \Delta {P_{{\rm W},t}} +  {C_{{\rm {outres}}}^t}
\end{array}
\end{equation}
where $[\cdot]_t$ represents the $t$ item. In addition, wind power utilization and $C_{\rm {ASC}}^{{\rm{UP}}}$ are also the main indicators on power dispatch. Then, the maximization of cooperative profit ${\psi _t}(x,y)$ is established as the optimal profit constraint to characterize profit-driven response. Thus, the model of intraday PDSCR is formulated as the multi-objectives optimization in (\ref{eq23}). The supplemental constraints of the thermal power and grid are shown in (B.16)-(B.19) of Appendix B.
\begin{equation}\label{eq23}
\begin{array}{l}
\begin{aligned}
\min \quad & [{J_5},{J_6}]\\[0.8mm]
{\rm {s.t.}}\quad &{J_5} = \sum\limits_{t = 1}^{\rm T} {P_{{\rm{WP,cur}}}^t} \\[0.8mm]
&{J_6} = 1 - C_{\rm {ASC}}^{{\rm{UP}}}\\[0.8mm]
&\arg \max \quad {\psi _t}(x,y)\\[0.8mm]
&(1), (6)- (10), \rm{(B.16)-(B.19)}
\end{aligned}
\end{array}
\end{equation}

\subsection{Multi-individual Profit Distribution Marginal Solution} 

{Regarding wind power fluctuations, the intraday PDSCR is devoted to balancing power supply and demand, while also alleviating the issue of branch overload. Correspondingly, the cooperative mechanism allows the demand-supply side to receive financial incentives. However, the implementation of PDSCR introduces cost variations in thermal power generation and industrial production. If the financial compensation fails to benefit individuals, or if the profit distributions are unreasonable, it would lead to the conflict of interests. In such cases, achieving effective autonomous cooperation becomes challenging. In the case study of Section V, we will test the performance of non-cooperation scenarios, in order to quantify the extent of negative impacts caused by conflicts.

\begin{figure}[htb]
	\vspace{-0.2cm}
	\centering
	\includegraphics[scale=0.34]{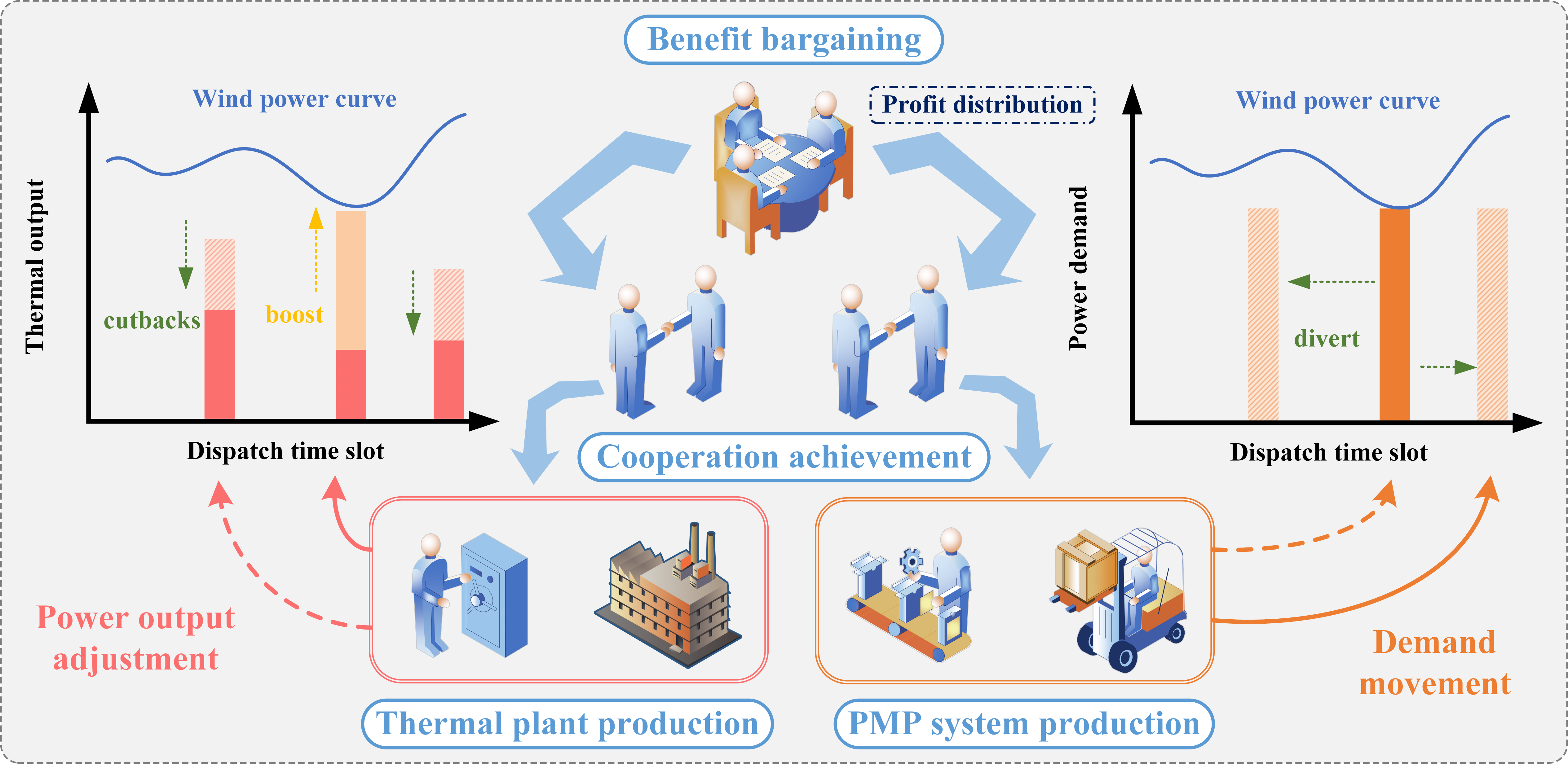}
	\vspace{0.2cm}
	\caption{Workflow of bargaining and coordination.}\label{Fig8}
	\vspace{-0.2cm}
\end{figure}

Note that the Nash bargaining theory is an effective framework for cooperation and conflicts of interest \cite{ref21}. As shown in Fig. \ref{Fig8}, it assumes there exists a bargaining process where multiple individuals discuss profit distributions. Each individual subsequently implements their response based on the outcome of these negotiations. However, it is worth noting that this theory primarily focuses on bargaining balanced solutions within the context of two-party linear utilities. In this subsection, we extend an optimal bargaining model to address the issue of multi-individual nonlinear utilities, that is, the Multi-Individual Optimal Marginal Model (MIOMM). Initially, we demonstrate the convexity and feasibility of MIOMM. Subsequently, we propose and derive Multi-Individual Profit Distribution Marginal Solutions (MIPDMSs) based on different distribution criteria, which in turn serve as multi-individual response constraints.}

{Note that the problem has the following characteristics: {\emph{(i)}} The potential conflicts of interests exist among individuals, i.e., the profit distribution of IPE and thermal plants. {\emph{(ii)}} As shown in (\ref{eq19}), the distribution of total profit is directly proportional to the increase of wind power utilization. Then, we propose the equal and contribution-based profit distributions. In this case, the more distribution means the more profit for each individual. For demand and supply sides, this incentive is the profit motivation for improving wind power utilization. {\emph{(iii)}} The behavior is not mandatory, it means each individual can refuse to respond. To this end, the profit utility $u_k$ could be established in the scenario of multi-individual negotiation \cite{ref22}, which is presented as follows.}
\begin{equation}\label{eq24}
\begin{array}{l}
S = \{ ({u_1},{u_2},...,{u_n})|{u_k} = {s_k}(b) , b \in B\} \\[1.5mm]
D = ({s_1}(d_1),{s_2}(d_2),...,{s_n}(d_n)) \\[1.5mm]
k = 1,2,...,n,
\end{array}
\end{equation}
where $S$ and $D$ denote the cooperative and divergent sets. ${s_k}(b)$ is the behavioral profit function for individual $k$, $b$ is the cooperative behavior and ${d_n}$ represents the disagreeable behavior. Here, the benefits pair $(S,D)$ provides individuals with the options of cooperation and non-cooperation. In the cooperative process, Nash has provided a linear equilibrium model called Nash bargaining solution (NBS) \cite{ref23}. In PDSCR, we expend this equilibrium to multi-interest and nonlinear profit utility, and this multi-individual model is formulated as (\ref{eq25}).
\begin{equation}\label{eq25}
\begin{array}{l}
\begin{aligned}
\max \quad &(V - {V_0})\prod\limits_{k = 1}^{N_{\rm{T}}} {({U_k} - U_k^0)} \\[1mm]
{\rm{s.t.}} \quad & {U_k} - U_k^0 \ge 0,\; V - {V_0} \ge 0
\end{aligned}
\end{array}
\end{equation}
where $U_k$ and $V$ represent individual costs of thermal plant $k$ and IPE, respectively. Their original costs are $U_k^0$, $V_0$. Then, profit utility functions of are defined as ${u_k} = {{U_k} - U_k^0}$, $v = V - {V_0}$, $({u_k},v) \in S$. In (\ref{eq25}), the profit non-inferiority ensures the motivation of individual participation. In intraday PDSCR, the specific profit utility function are redefined as (\ref{eq26}), where $U_k$, $V$, $U_k^0$ and $V_0$ are established in independent period ${U_{t,k}}$, $U_{t,k}^0$, ${V_t}$ and $V_t^0$. There is a quadratic relationship between the multi-individual profit function and the related decision variable.
\begin{equation}\label{eq26}
\begin{array}{l}
\begin{aligned}
{u_{k,t}} & = {U_{t,k}} - U_{t,k}^0 = \sum\limits_{p \in k} {[F({P_{{\rm {P}},t,p}}) - F(P_{{\rm {P}},t,p}^{{\rm {UP}}})} ] - S_k^V \cdot \\
&({v_t} - S_t^{{\rm {P,R}}}) - C_{{\rm outres}}^t + S_{{\rm T},t,k}^{\rm R} \\
{v_t} &= {V_t} - V_t^0 = {F_{{\rm P},t}} + {F_{{\rm B},t}} - F_{{\rm P},t}^{\rm UP} - F_{{\rm B},t}^{\rm UP} + S_t^{{\rm P,R}}
\end{aligned}
\end{array}
\end{equation}
where $S_k^V$ denotes the income coefficient, which indicates the increased electricity sales revenue in thermal plant $k$. The total profit ${\psi _t}(x,y)$ can be further reformulated as follows. 
\begin{small}
\begin{equation}\label{eq27}
\begin{array}{l}
\begin{aligned}
\max \quad {\psi _t}(x,y) = &\sum\limits_{k = 1}^{N_{\rm{T}}} {{u_{k,t}}}  + {v_t} = \sum\limits_{i = 1}^{{N_T}} {({U_{t,k}} - U_{t,k}^0)}  + ({V_t} - V_t^0)
\end{aligned}
\end{array}
\end{equation}
\end{small}

Note that the set $({u_{k,t}},v_t)$ of profit utility functions is multidimensional and quadratic. In the PDSCR strategy, the convexity of $({u_{k,t}},v_t)$ is the necessary and sufficient condition of MIOMM optimality. Therefore, we derive the convexity of the profit utility, and the lemma and proof are shown as follows.

{\emph{\textbf{Lemma 1:}}} For $(u,v)$, $u\in {S_1} : \{ u = {x^ \top }Ax + {c_1}^ \top x|a_1^ \top x \le {b_1}, a_2^ \top x = {b_2},u \in {\mathbb{R}^n}\}, v\in {S_2} : \{ v = {c_2}^ \top y|a_3^ \top y \le {b_3},a_3^ \top y = {b_3},y \in \mathbb{R}{^m}\} $, set $S : \{ (u,v)|{u_i},{v_j} > 0,u \in \mathbb{R}{^n},v \in \mathbb{R}{^m}\} $ is convexity, where $A = diag\{ {a_1},{a_2},...,{a_n}\} $.

{\emph{\textbf{Proof:}}} The detailed proof could be referred to Appendix C.

$\hfill\blacksquare$

In MIOMM, we further study the optimality of cooperative total profit ${\psi _t}(x,y)$, which is to prove the equivalent optimality in both models. On this basis, we could replace $\max({\psi _t}(x,y))$ with MIOMM.

{\emph{\textbf{Proposition:}}} The maximization of multiplication $\prod\limits_{{\chi _m} \in \chi } {{\chi _m}}$ achieves the maximization of accumulation $\sum\limits_{{\chi _m} \in \chi } {{\chi _m}}$ for ${\chi _m}$.

{\emph{\textbf{Proof:}}} The multiplication and accumulation are defined as ${f_\Pi }$ and ${f_\Sigma }$ ,
\begin{equation}\label{eq28}
\begin{array}{l}
\begin{aligned}
\max \quad &{f_\Pi } = \prod\limits_{{\chi _m} \in \chi } {{\chi _m}}  ,{f_\Sigma } = \sum\limits_{{\chi _m} \in \chi } {{\chi _m}} \\
{\rm {s.t.}}\quad &{\chi _m} \ge 0.
\end{aligned}
\end{array}
\end{equation}

In the linear function ${f_\Sigma }(\chi )$, $d$ denotes the direction vector and $\forall {d_m} \in d$, ${d_m} > 0$. Obviously, ${f_\Sigma }(\chi  + d) > {f_\Sigma }(\chi )$, i.e., $d$ is able to represent the optimal direction of ${f_\Sigma }$. For multiply ${f_\Pi }(\chi  + d)$, the first-order Taylor unfolding is as (\ref{eq29}) when ${\chi _m} \ge 0$.
\begin{equation}\label{eq29}
{f_\Pi }(\chi  + d) = {f_\Pi }(\chi ) + \nabla {f_\Pi }{(\chi )^ \top }d + o(d)
\end{equation}
where $o(d) > 0$. When ${d_m} > 0$, the partial derivative $\nabla {f_\Pi }{(\chi )^ \top }d$ can be expressed as (\ref{eq30}). 
\begin{equation}\label{eq30}
\nabla {f_\Pi }{(\chi )^ \top }d = \sum\limits_{{d_m} \in d} {(\prod\limits_{\scriptstyle{\chi _r} \ne {\chi _m},\hfill\atop
\scriptstyle{\chi _r} \in \chi \hfill} {{\chi _r}} ) \cdot {d_m}}  > 0.
\end{equation}

Therefore, ${f_\Pi }(\chi  + d) > {f_\Pi }(\chi )$, i.e., the maximization of accumulation ${f_\Sigma }$ is ensured by maximizing the multiplication ${f_\Pi }$.

$\hfill\blacksquare$

On the basis of MIOMM, the optimal marginal solution is derived in the convex feasible region called MIPDMS. We assume that multi-individuals would require to divide the cooperative profits, according to the equal or contribution-based criteria. It could respectively represent the fairness and rationality in cooperation, which are the responding motivations for multi-individuals. In the equal MIPDMS, this criteria provide the same profit to participants. In contribution-based MIPDMS, profit utility is formulated according to their actual contributions in cooperation. Accordingly, both the above solutions are feasible, which are demonstrated as follows.

Firstly, the equal MIPDMS in the fair distribution condition is developed within the above multi-individual model, i.e., maximizing the multiplication of multi-individual utility.

{\emph{\textbf{Lemma 2:}}} For thermal plants and PMP, the profit is distributed by each individual ${u_{m,t}}$ or ${v_t}$ in equal MIPDMS as (\ref{eq31}).
\begin{equation}\label{eq31}
{u_{m,t}} = {v_t} = \frac{{{\psi _t}(x,y)}}{{{N_{\rm{T}}} + 1}}
\end{equation}

{\emph{\textbf{Proof:}}} For the NBS objective as (\ref{eq25}), the function could be shown as follows.
\begin{equation}\label{eq32}
\mathop {\min }\limits_{{u_{m,t}},{v_t}}   \ln \; {v_t} + \sum\limits_{k = 1}^{N_{\rm{T}}} {\ln } \; {u_{k,t}}
\end{equation}
where the composite function (19) is convex, the minimum will correspond to the primeval optimal solution. The partial derivative to $u_{m,t}$ is as (\ref{eq33}).
\begin{small}
\begin{equation}\label{eq33}
\begin{array}{l}
{\nabla _{{u_{m,t}}}}(\ln {v_t} + \sum\limits_{k = 1}^{N_{\rm{T}}} {\ln }  {u_{k,t}}) = ({{\sum\limits_{k = 1}^{N_{\rm{T}}} {{u_{k,t}}}  - \psi (x,y)}})^{-1} + ({{{u_{m,t}}}})^{-1}
\end{array}
\end{equation}
\end{small}

In the optimal point, ${\nabla _{{u_{m,t}}}}=0$, i.e. $\psi (x,y) = \sum\limits_{k = 1}^{N_{\rm{T}}} {{u_{k,t}}}  + {u_{m,t}}$. If the profit is distributed equally, we have the following formulations.
\begin{equation}\label{eq34}
{u_{m,t}} = \frac{{{\psi _t}(x,y)}}{{{N_{\rm{T}}} + 1}},\quad {v_t} = \psi (x,y) - \sum\limits_{k = 1}^{N_{\rm{T}}} {{u_{k,t}}}  = \frac{{{\psi _t}(x,y)}}{{{N_{\rm{T}}} + 1}}
\end{equation}

$\hfill\blacksquare$

On the other hand, we consider the rational distribution condition to derive the contribution-based MIPDMS. In this solution, the more responding contribution of individuals means the more proportion of total profit.

{\emph{\textbf{Lemma 3:}}} For thermal plants and PMP, the profit is divided by each individual in contribution-based MIPDMS as (\ref{eq35}-\ref{eq36}).
\begin{equation}\label{eq35}
{u_{m,t}} = {\sigma _{m,t}} \cdot {\psi _t}(x,y)/(\sum\limits_{k = 1}^n {{\sigma _{k,t}}}  + {\sigma _{m,t}})
\end{equation}
\begin{equation}\label{eq36}
v_t = {\psi _t}(x,y) \cdot [\sum\limits_{k = 1}^{N_{\rm{T}}} {{\sigma _{k,t}}}  \cdot \sum\limits_{j = 1}^{N_{\rm{T}}} {{{(\sum\limits_{k = 1}^{N_{\rm{T}}} {{\sigma _{k,t}}}  + {\sigma _{j,t}}}}})^{ - 1}  - {N_{\rm{T}}} + 1]
\end{equation}
where ${\sigma _{k,t}} = \sum\limits_{p \in k}^{{N_T}} {({P_{T,p,t}} - P_{T,p,t}^{UP})}$ represents the ranger of adjusted power of the $k^{th}$ thermal plant.

{\emph{\textbf{Proof:}}} Based on (\ref{eq33}), we obtain the following formulation.  
\begin{equation}\label{eq37}
\psi (x,y) = \sum\limits_{k = 1}^{N_{\rm{T}}} {{u_{k,t}}}  + {u_{m,t}}
\end{equation}

When the profit distribution is based on the adjusted weights, the total profit function is reformulated as (\ref{eq38}).
\begin{equation}\label{eq38}
\psi (x,y) = \sum\limits_{k = 1}^{N_{\rm{T}}} {\frac{{{\sigma _k}}}{{{\sigma _m}}}{u_{m,t}}}  + {u_{m,t}}
\end{equation}
where ${{{{\sigma _{k,t}}}}/{{{\sigma _{m,t}}}}}$ is defined as corresponding weights between ${{\sigma _{m,t}}}$ and ${{\sigma _{k,t}}}$.
\begin{equation}\label{eq39}
\begin{array}{l}
{u_{m,t}} = {\sigma _{m,t}} \cdot \psi (x,y)/(\sum\limits_{k = 1}^{N_{\rm{T}}} {{\sigma _{k,t}}}  + {\sigma _{m,t}}),\\
{v_t} = \psi (x,y) \cdot \{ 1 - \sum\limits_{j = 1}^{N_{\rm{T}}} {[{\sigma _{j,t}}/(\sum\limits_{k = 1}^{N_{\rm{T}}} {{\sigma _{k,t}}}  + {\sigma _{j,t}})]\} } \\
 = \psi (x,y) \cdot [\sum\limits_{k = 1}^{N_{\rm{T}}} {{\sigma _{k,t}}}  \cdot \sum\limits_{j = 1}^{N_{\rm{T}}} {{{(\sum\limits_{k = 1}^{N_{\rm{T}}} {{\sigma _{k,t}}}  + {\sigma _{j,t}}}}})^{ - 1}  - {N_{\rm{T}}} + 1]
\end{array}
\end{equation}

$\hfill\blacksquare$

On the basis of the contribution-based MIPDMS, we explore the specific individual profit boundary as follows.

{\emph{\textbf{Lemma 4:}}} For IPE profit $v_t$ as (\ref{eq36}), upper boundary: $v_t \le {\rm max}[u_{k,t}]$, $i = 1,2,...,n$, distribution proportion boundary: $v_t \in [\psi (x,y)/(n+1),\psi (x,y)/2]$.

{\emph{\textbf{Proof:}}} Firstly, $u_{m,t} = {\rm max}[u_{k,t}]$, $\sigma_{m,t} = {\rm max}[\sigma_{k,t}]$, and the relationship of $u_{k,t}$ and $v_{t}$ can be represented as follows.
\begin{equation}\label{eq41}
\begin{array}{l}
\begin{aligned}
&{{({u_{m,t}} - {v_t})}}/{{\psi (x,y)}} = \\
&  {\sigma _{m,t}}/(\sum\limits_{k = 1}^{N_{\rm{T}}} {{\sigma _{k,t}}}  + {\sigma _{m,t}}) + \sum\limits_{j = 1}^{N_{\rm{T}}} {[{\sigma _{j,t}}/(\sum\limits_{k = 1}^{N_{\rm{T}}} {{\sigma _{k,t}}}  + {\sigma _{j,t}})]}  - 1 = \\
&  \sum\limits_{j = 1}^{N_{\rm{T}}} {[{\sigma _{j,t}}/(\sum\limits_{k = 1}^{N_{\rm{T}}} {{\sigma _{k,t}}}  + {\sigma _{j,t}})]}  - \sum\limits_{k = 1}^{N_{\rm{T}}} {{\sigma _{k,t}}} /(\sum\limits_{k = 1}^{N_{\rm{T}}} {{\sigma _{k,t}}}  + {\sigma _{m,t}}) \ge 0.
\end{aligned}
\end{array}
\end{equation}

Therefore, ${\rm max}[u_{k,t}]$ is the upper boundary of IPE profit $v_t$. For the IPE profit $v_t$, 
${v_t} = \{ 1 - \sum\limits_{j = 1}^{N_{\rm{T}}} {[{\sigma _{j,t}}/(\sum\limits_{k = 1}^{N_{\rm{T}}} {{\sigma _{k,t}}}  + {\sigma _{j,t}})]} \}  \cdot \psi (x,y)$, its 1{\emph{st}} and 2{\emph{nd}} partial derivatives are shown as follows.
 
\begin{equation}\label{eq42}
\begin{array}{l}
\begin{aligned}
{\nabla _{{\sigma _{m,t}}}}{v_t} = &\sum\limits_{j = 1}^{N_{\rm{T}}} {[{\sigma _{j,t}}/{{(\sum\limits_{k = 1}^{N_{\rm{T}}} {{\sigma _{k,t}}}  + {\sigma _{j,t}}}})^2]} - \sum\limits_{k = 1}^{N_{\rm{T}}} {{\sigma _{k,t}}} / \\ 
&{(\sum\limits_{k = 1}^{N_{\rm{T}}} {{\sigma _{k,t}}}  + {\sigma _{m,t}}})^2
\end{aligned}
\end{array}
\end{equation}
\begin{equation}\label{eq43}
\begin{array}{l}
\begin{aligned}
\nabla _{{\sigma _{m,t}}}^2{v_t} = &(4 \cdot \sum\limits_{k = 1}^{N_{\rm{T}}} {{\sigma _{k,t}}}  - 2 \cdot {\sigma _{m,t}})/{(\sum\limits_{k = 1}^{N_{\rm{T}}} {{\sigma _{k,t}}}  + {\sigma _{m,t}})^3} - \\
& 2 \cdot \sum\limits_{j = 1}^{N_{\rm{T}}} {[{\sigma _{j,t}}/{{(\sum\limits_{k = 1}^{N_{\rm{T}}} {{\sigma _{k,t}}}  + {\sigma _{j,t}}}})^3]} 
\end{aligned}
\end{array}
\end{equation}

There are two types of extreme points in feasible domain $\sigma _{m,t} \in [0,\sum\limits_{k = 1}^{N_{\rm{T}}} {{\sigma _{k,t}}}]$. ({\emph {i}) For ${\sigma _{m,t}} \in \{ {\sigma _{k,t}}|{\sigma _{1,t}} = {\sigma _{2,t}} = ... = {\sigma _{{N_{\rm{T}}},t}},k \in 1,2,...,{N_{\rm{T}}}\}$, $\nabla _{{\sigma _{m,t}}}^2{v_{m,t}} = \frac{{2(n - 1) \cdot {\sigma _{m,t}}}}{{{{[({N_{\rm{T}}} + 1) \cdot {\sigma _{m,t}}]}^3}}} > 0$, and $\sigma _{m,t}$ corresponds the local lower boundary. For $\forall {\sigma _{m,t}} \in [0,\sum\limits_{k = 1}^{N_{\rm{T}}} {{\sigma _{k,t}}} /{N_{\rm{T}}}]$ and $\forall {\sigma _{m,t}} \in (\sum\limits_{k = 1}^{N_{\rm{T}}} {{\sigma _{k,t}}} /{N_{\rm{T}}},\sum\limits_{k = 1}^{N_{\rm{T}}} {{\sigma _{k,t}}} ]$, which constitutes the complete feasible domain, ${\nabla _{{\sigma _{m,t}}}}{v_{m,t}} \le 0$ and ${\nabla _{{\sigma _{m,t}}}}{v_{m,t}} \ge 0$ separately, and $\sigma _{m,t}$ corresponds the global lower boundary, where $v_t = \frac{\psi (x,y)}{({N_{\rm{T}}}+1)} $. ({\emph {ii}) For ${\sigma _{m,t}} \in \{ {\sigma _{k,t}}|{\sigma _{m,t}} = \sum\limits_{k = 1}^{N_{\rm{T}}} {{\sigma _{k,t}}} ,{\sigma _{k,t}} = 0,i \ne m,k = 1,2,...,{N_{\rm{T}}}\} $, $\nabla _{{\sigma _{m,t}}}^2{v_{m,t}} = 0$, $\sigma _{m,t}$ corresponds the global upper boundary, where $v_t = \frac{\psi (x,y)}{2} $.

$\hfill\blacksquare$

Accordingly, the total profit optimality constraint could be replaced by MIPDMS, and the intraday PDSCR model is reformulated as (\ref{eq41}).
\begin{equation}\label{eq41}
\begin{array}{l}
\begin{aligned}
\min \quad &[{J_5},{J_6}]\\[1mm]
{\rm{s.t.}} \quad &{J_5} = \sum\limits_{t = 1}^T {P_{{\rm{WP,cur}}}^t} \\[1mm]
&{J_6} = 1 - C_{\rm {ASC}}^{{\rm{UP}}}\\[1mm]
&{\rm (10),(18),(22)-(23)}
\end{aligned}
\end{array}
\end{equation}

It is worth mentioning that the MIPDMS in different criteria stands for the optimal profit utility sets. In the above two forms of MIPDMS, the demand-supply cooperative response is satisfactory in the profit distribution and is without the conflict of interest. The reason for this is that multi-individual profits are maximized, and the corresponding distribution is fair and rational. Furthermore, specific numerical relationships are analyzed in Section V.


\section{Case study}
\subsection{Data Preparations}

A modified IEEE 24-bus power system is used to conduct case studies, in which two wind farms are located on bus 12 and bus 9, and a PMP system representing an IPE is connected to bus 10. Moreover, we set 6 procedures and 5 buffers in the PMP system, where the maximum projects of $M_i$ and $B_i$ are 2 and 4. The powers of the unit projects in procedure and buffer are as follows: $\{96,64,24,72,64,42\}$ and $\{8,8,8,8,8,8\}$. The minimum production quantity of the PMP system in a dispatch cycle is set as 25, and information on DR programs is shown in TABLE II, where $\Delta P_{\rm W}$ represents the increase in wind power utilization.

\begin{table}[h]\scriptsize
   \renewcommand{\arraystretch}{1.5}
	\begin{center}
		\caption{PARAMETERS OF DR PROGRAMS}
		\begin{tabular}{m{1.0cm}<{\centering}m{1.2cm}<{\centering}m{1.2cm}<{\centering}m{2.3cm}<{\centering}m{0.9cm}<{\centering}} \hline\hline	
			DR program &  Demand condition  & Demand periods(h)  & Consumption price ($\$$/kWh)  & Fixed cost ($\$$)  \\   
			\hline
			Price-based DR      & WP trend & 1-24      & 0.16790/0.08274   & 51.42   \\          
			Incentive-based DR   & WP fluctuation      & 1-24    & -0.15$\cdot \Delta P_{\rm W}$           & 0.00   \\   
			\hline\hline	
		\end{tabular}
	\end{center}
	\vspace{-0.3cm}
\end{table}

In this paper, three cases are conducted to verify the effectiveness of the PDSCR strategy, while revealing impacts of the equal and the contribution-based MIPDMSs on the performance of demand-supply response. As shown in TABLE III, Case 1 is based on the day-ahead PDSCR strategy, regarded as a benchmark (without cooperation). By comparing Case 1 with Cases 2$\sim$3, we could verify the optimization effectiveness of the intraday PDSCR strategy for wind power utilization and transmission margin enhancement. {The difference between Cases 2 and 3 is that the cooperation model respectively implements the equal and the contribution-based MIDPMSs. It means that the cooperative profit is divided equally or not. Therefore, by comparing Case 2 with Case 3, we could analyze the impacts of these two MIPDMSs.}
\vspace{-0.3cm}
\begin{table}[h]\scriptsize
   \renewcommand{\arraystretch}{1.3}
	\begin{center}
		\caption{CASES STUDIES IN STEP 2 UNCERTAIN SCENARIO}
		\begin{tabular}{m{1.0cm}<{\centering}m{2.5cm}<{\centering}m{1.3cm}<{\centering}m{2.3cm}<{\centering}} \hline\hline	
			Case &  Strategy  & Constraints   & Gaussion $\sigma$ of scenario  \\   
			\hline
			Case 1   & Day-ahead strategy     & —        & 0.08   \\   
			Case 2   & Intraday strategy 1    & (31)      & 0.08   \\          
			Case 3   & Intraday strategy 2    & (35-36)   & 0.08   \\ 
			\hline\hline	
		\end{tabular}
	\end{center}
	\vspace{-0.4cm}
\end{table}

In order to characterize wind power uncertainty in the intraday scheme, the intraday wind power error is assumed to follow the Gaussian distribution. In this case, the forecast value is considered as the mean and the standard deviation is set to 0.08 of the mean value. Based on this distribution model, the {\emph {Latin hypercube}} approach \cite{ref24}, a traditional method, is used to sample the wind power values. Then, the sampling values constitute the multiple random scenarios to characterize wind power fluctuation, and the three cases are simulated with 100 scenarios.

\subsection{Analysis of Day-ahead PDSCR}

In this subsection, we obtain the scheme of Case 1 by the day-ahead PDSCR strategy. The bi-objective Pareto front (PF) is shown as Fig. \ref{Fig10}, which is the curve of objective values regarding obtained Pareto solutions. It includes the two upper-level objectives of $C_{\rm{ASC}}$ (optimal boundary: 0.7728) and total thermal cost (optimal boundary: 120652.49$\$$), as well as corresponding wind power curtailments. PF well characterize the trade-off relationship between the above two objectives. In Section III, we illustrate the consistent interests between the objectives of wind power curtailment and thermal cost. It is validated by the wind power curtailment curve in Fig. \ref{Fig10}, where the wind power curtailment decreases with the reduction of thermal cost.

\begin{figure}[htb]
	\vspace{-0.2cm}
	\centering
	\includegraphics[scale=0.223]{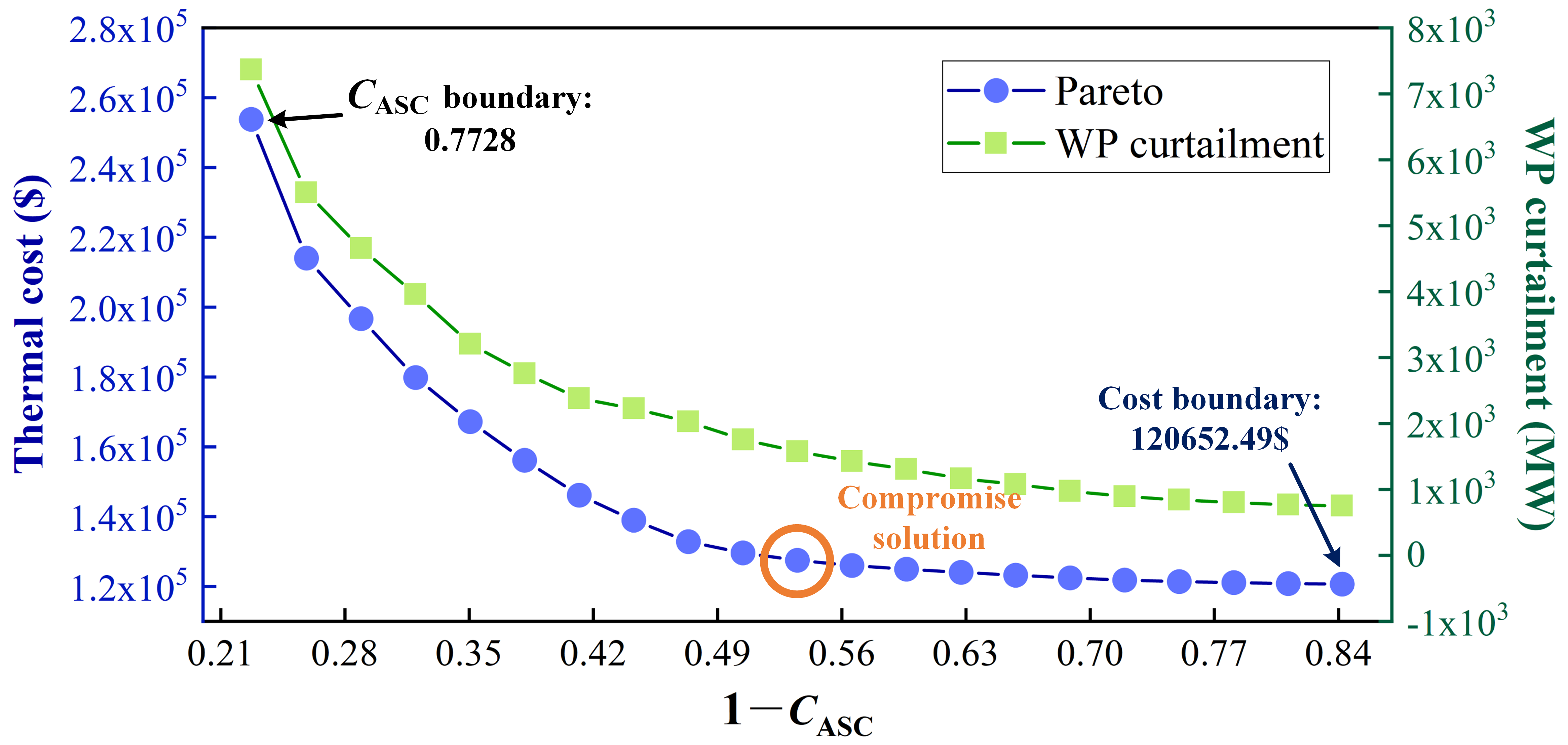}
	\vspace{-0.3cm}
	\caption{Pareto front and corrsponding WP curtailment.}\label{Fig10}
	\vspace{-0.4cm}
\end{figure}

Also, it is necessary to verify the rationality of the determined day-ahead PDSCR scheme. Herein, we analyze the PF compromise solution in Fig. \ref{Fig10}, which is the center point in the $C_{\rm{ASC}}$ boundary. Specifically, the responding curve of demand and thermal power are shown in Fig. \ref{Fig11}(a). During the peak periods of wind power supply, such as 8:00-11:00 and 17:00-19:00, the total power demand shows an upward trend, when thermal power is relatively reduced. In contrast, the total power demand decreases during the valley periods of wind power supply, such as 1:00-6:00 and 12:00-16:00. This tracking trend minimizes the thermal power and the wind power curtailment, and reveals that the shifting of industrial power demand and the thermal operation scheme are matching the operation demand of power system in day-ahead dispatch.

\begin{figure}[htb]
	\vspace{-0.2cm}
	\centering
	\includegraphics[scale=0.263]{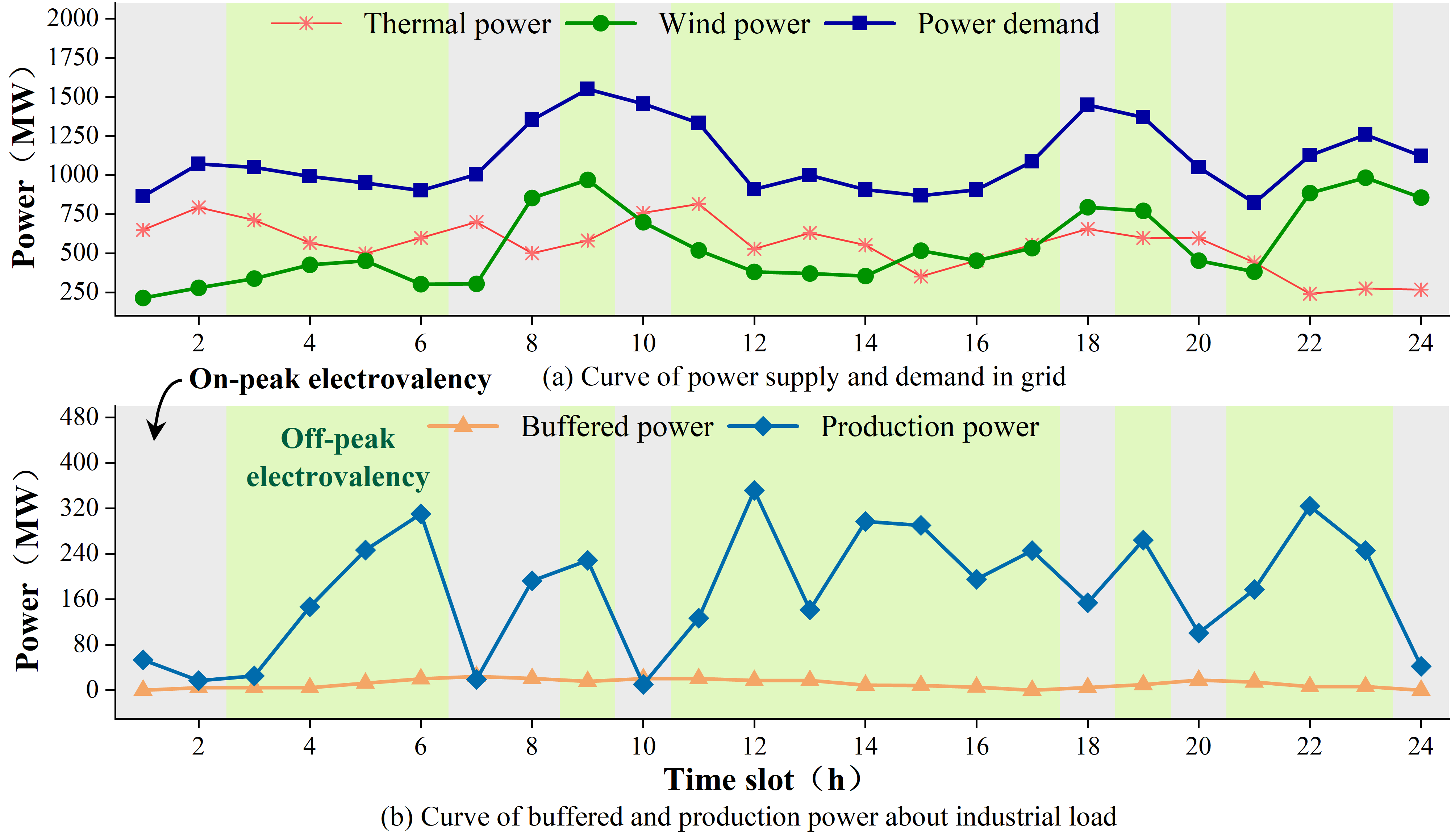}
	\vspace{-0.2cm}
	\caption{Dispatching result in day-ahead.}\label{Fig11}
	\vspace{-0.0cm}
\end{figure}

\begin{figure}[!h]
	\vspace{-0.2cm}
	\centering
	\includegraphics[scale=0.237]{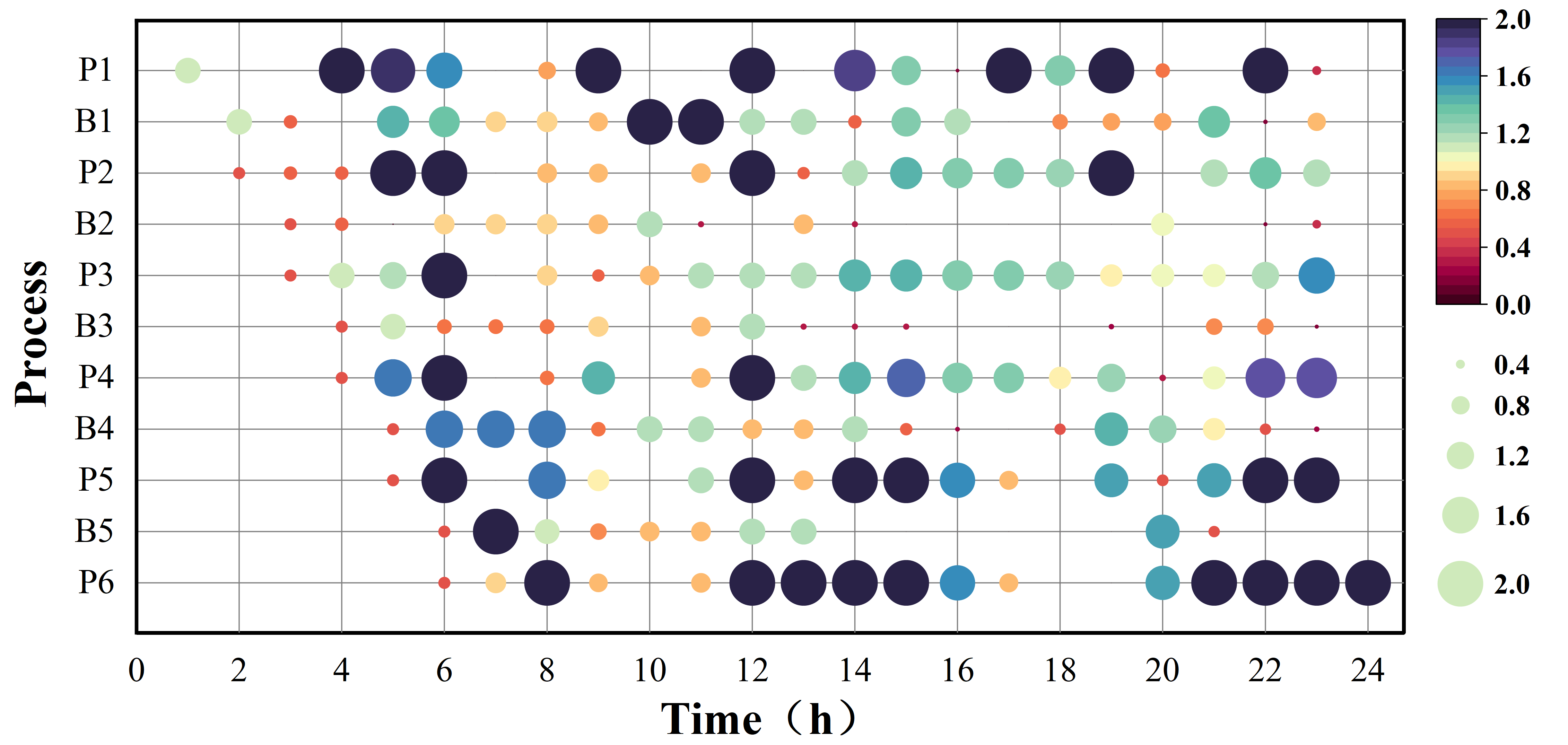}
	\vspace{-0.2cm}
	\caption{Details of PMP production dispatch.}\label{Fig12}
	\vspace{-0.2cm}
\end{figure}

\begin{table*}[!h]\scriptsize
   \renewcommand{\arraystretch}{1.2}
	\begin{center}
	\renewcommand{\multirowsetup}{\centering}
		\caption{Simulation results in probabilistic sampling scenarios}
		\begin{tabular}{m{0.8cm}<{\centering}|m{1.7cm}<{\centering}| m{0.8cm}<{\centering}m{0.8cm}<{\centering}m{0.8cm}<{\centering}|m{0.9cm}<{\centering}m{0.9cm}<{\centering}m{0.9cm}<{\centering}|m{0.8cm}<{\centering}m{0.8cm}<{\centering}m{0.8cm}<{\centering}|m{0.9cm}<{\centering}m{0.9cm}<{\centering}} \hline\hline	
		{\multirow{2}{*}{Scenario}} & {\multirow{2}{1.9cm}{WP fluctuation /$10^3\cdot$MW}} & \multicolumn{3}{c|}{WP curtailment /$10^3\cdot$MW} & \multicolumn{3}{c|}{Transmission margin $\rm C_{ASC}/10^{-2}$} & \multicolumn{3}{c|}{Total cost /$10^5\cdot\$$} & \multicolumn{2}{c}{Hypervolume /$10^{-2}$} \\ \cline{3-13}
 &  & Case 1 & Case 2 & Case 3  & Case 1 & Case 2 & Case 3  & Case 1 & Case 2 & Case 3 & Case 2  & Case 3\\ \hline

	1   & 13.15  & 14.58   & 10.01   & 10.46   & 38.93  & 37.08  & 32.53  & 6.06   & \textbf {4.69}   & 4.80  & \textbf {14.87}   & 13.76   \\          
	2   & 3.96   & 5.39    & 2.39   & 2.42   & 38.93  & 31.90  & 31.42  & 6.06   & \textbf {5.03}   & 5.09   & \textbf {37.39}   & 36.42   \\ 
	3   & 2.55   & 4.27    & 1.79   & 1.79   & 38.93  & 37.17  & 36.72  & 6.12   & \textbf {5.20}   & 5.21   & \textbf {39.54}   & 38.55   \\ 
	4   & 1.58   & 3.01    & 1.44   & 1.43   & 38.93  & 33.76  & 33.30  & 6.06   & 5.31   & \textbf {5.30}   & \textbf {40.40}   & 39.43   \\
	5   & 1.10   & 3.07    & 1.26   & 1.25   & 38.93  & 35.90  & 35.42  & 6.17   & \textbf {5.31}   & 5.33   & \textbf {40.99}   & 40.02   \\
	6   & 0.48   & 3.11    & 0.98   & 0.98   & 38.93  & 37.10  & 36.77  & 6.31   & 5.31   & \textbf {5.24}   & \textbf {41.81}   & 40.79   \\
	7   & -0.35  & 1.52    & 0.75   & 0.78   & 38.93  & 33.24  & 34.21  & 6.15   & 5.07   & \textbf {5.00}   & \textbf {47.13}   & 46.58   \\
	8   & -1.63  & 1.22    & 0.41   & 0.42   & 32.91  & 30.88  & 30.88  & 6.35   & 5.19   & \textbf {5.07}   & \textbf {47.87}   & 46.99   \\
	9   & -2.57  & 0.77    & 0.15   & 0.15   & 33.76  & 35.37  & 34.45  & 6.45   & 5.30   & \textbf {5.13}   & \textbf {48.39}   & 47.49   \\
	10   & -3.63  & 0.60   & 0.06   & 0.06   & 30.46  & 39.74  & 38.78  & 6.62   & 5.38   & \textbf {5.32}   & \textbf {47.96}   & 47.00   \\ 
	11   & -5.14  & 0.41   & 0.03   & 0.05   & 25.41  & 43.36  & 39.03  & 6.89   & \textbf {5.40}   & 5.45   & \textbf {43.66}   & 39.41   \\ 
			\hline\hline	
		\end{tabular}
	\end{center}
	\vspace{-0.7cm}
\end{table*}

Specifically, the corresponding power demand trend is influenced by price-based IDR, and the specific industrial demand curve is illustrated in Fig. \ref{Fig11}(b). When the utility center determines low electricity prices (green areas) during certain periods, such as 3:00-6:00 or 12:00-17:00, industrial demand increases. During these periods, IPEs initiate projects in the $P$ process, as depicted by P1-P6 in Fig. \ref{Fig12}, to maximize production. The color depth and size of these projects are directly proportional to the number of ongoing projects. Conversely, when the utility center determines high electricity prices (gray areas) at 7:00 and 10:00, IPE initiates projects in the $B$ process, denoted as B1-B5 in Fig. \ref{Fig12}, resulting in a reduction in industrial demand during these periods.

\begin{figure*}[!b]
	\vspace{-0.2cm}
	\centering
	\includegraphics[scale=0.22]{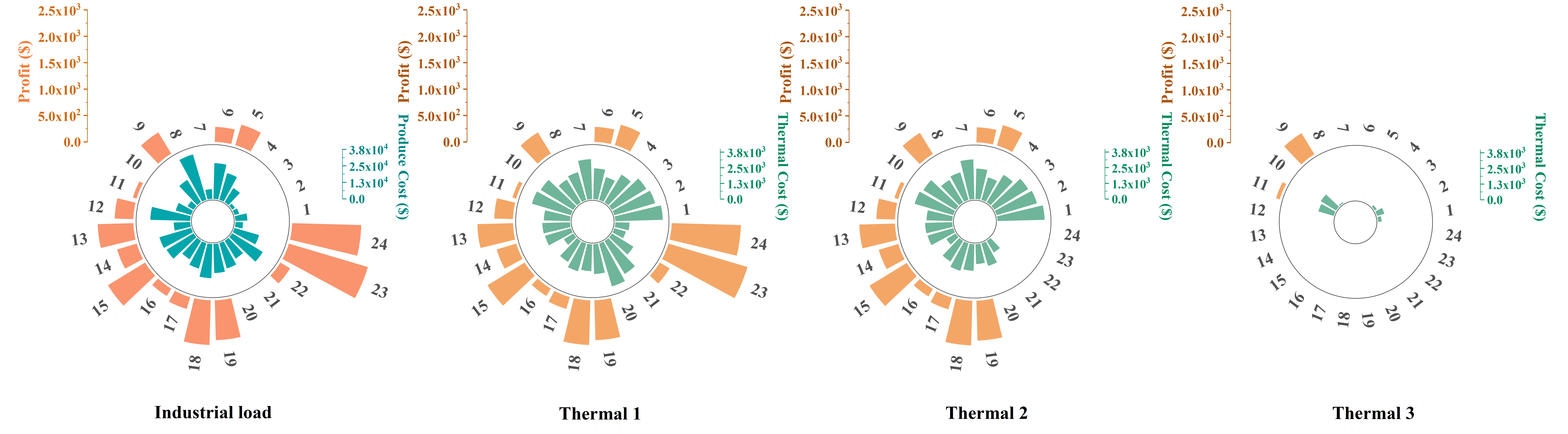}
	\vspace{0cm}
	\caption{Demand-supply cost and profit changes in Case 2.}\label{Fig13}
	\vspace{-0.1cm}
\end{figure*}

\begin{figure*}[!b]
	\vspace{-0.2cm}
	\centering
	\includegraphics[scale=0.22]{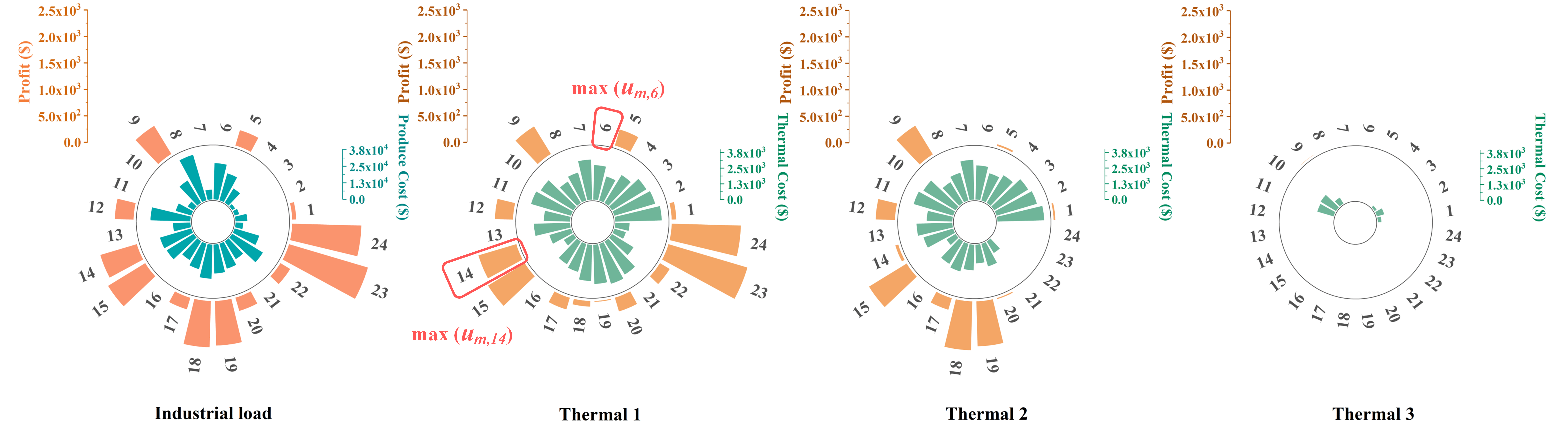}
	\vspace{0cm}
	\caption{Demand-supply cost and profit changes in Case 3.}\label{Fig14}
	\vspace{-0.1cm}
\end{figure*}

\subsection{Analysis of Intraday PDSCR}

{The day-ahead strategy is designed based on the standard wind power scenario without considering power fluctuations. In Case 1, we evaluate the performance of this day-ahead load scheme in uncertain scenarios. Cases 2 and 3, on the other hand, adapt the response scheme to accommodate uncertain scenarios using the intraday PDSCR strategy, where the equal and contribution-based MIPDMSs represent two different profit distribution approaches among cooperative participants. The analysis comprises two primary components: {\emph {i)}} Impact of different cases, and {\emph {ii)}} Cooperative profit distribution and detailed responding.}

{\emph {i)}} {\emph {Impact of different cases}}

In TABLE IV, we show the simulation results of 10 scenarios, which are selected in 100 random scenarios by stratified sampling based on the wind power fluctuation value. These results include the wind power curtailment, transmission margin $\rm C_{ASC}$, the total cost on the demand-supply side, and PF quality. Herein, hypervolume (HV) is an effective indicator to evaluate the PF quality \cite{ref25}. Specifically, compared with Case 1, intraday PDSCR strategies in Cases 2 and 3 achieve less wind power curtailment in all the representative scenarios. {They also exhibit a more stable $C_{\rm{ASC}}$, with a higher lower boundary (30.88) than that of Case 1 (0.2541). Furthermore, we observe that intraday PDSCR leads to a reduction in total costs, which results from optimizing total profits on the demand and supply side. Therefore, intraday strategies outperform day-ahead schemes in the conditions of wind power fluctuations, which ultimately benefits participating stakeholders on the demand-supply side.}

{Furthermore, by comparing Case 2 with Case 3, we can analyze which cooperative distribution criteria provide advantages. In this context, HV serves as an objective indicator, which is used to comprehensively assess the performance of multi-objective optimization. In TABLE IV, an examination of the HV values in Cases 2 and 3 reveals that the PF in the equal MIPDMS case outperforms the contribution-based one. Therefore, the PDSCR strategy that is constrained by equal MIPDMS can effectively enhance cooperative responses. }

{\emph {ii)}} {\emph {Cooperative profit distribution and detailed responding}}

The simulation results discussed above indicate that MIPDMSs are advantageous for increasing total profit and facilitating effective cooperation. In this part, we will study the specific profit distribution and the corresponding demand-supply responses within the two MIPDMS scenarios. There exists a positive linear relationship between the increment in wind power utilization $\Delta P_{\rm W}$ and the total profit function ${\psi _t}(x,y)$. When the positive fluctuation is small, profits tend to be relatively limited, potentially posing challenges in promoting cooperation. Hence, this paper selects Scenario 5 (fluctuation: 1100MW), where wind power is close to the forecast value, to explore the characteristics of MIPDMS.

In Cases 2 and 3, profit distributions among the participating individuals are illustrated in Fig. \ref{Fig13} and Fig. \ref{Fig14}. In these figures, the inner circle represents the final production cost after the response in each time period, while the outer circle represents the responding profit, which is the cost reduction. Note that some individuals are not in operation and do not receive a distribution of cooperative profit, e.g., for the thermal plant 3 during 12:00-24:00. 

In Case 2, we set the issue that the equal distribution of responding profits is required by multiple individuals on the demand-supply side. Consequently, in Fig. \ref{Fig13}, the multi-individual responding profit is equal during the cooperative periods, such as 11:00-19:00. When no total profit is obtained, individual responding profit is also 0, as observed during 1:00-4:00. This aligns with the equal MIPDMS constraint. In Case 3, we set that the contribution-based distribution of responding profit is required by multiple individuals. Consequently, Fig. \ref{Fig14} shows that cooperative profits vary among individuals. Taking 18:00-19:00 as an example, thermal plant 2 achieves higher profits, compared with thermal plant 1. It indicates that plant 2 has made a more contribution during these periods. These observations reveal that MIPDMS accurately distributes profits, and the fair profit distribution helps mitigate the conflict of interests among multiple individuals.

Moreover, to analyze the specific response scenarios in both cases, we present the load and power curves for demand-supply response in Fig. \ref{Fig15} and Fig. \ref{Fig16}. These figures also include the wind power consumption curve and the total profit. It can be seen that the reduction in thermal power supply and the increase in industrial load are advantageous for utilizing wind power. Conversely, in Fig. \ref{Fig16}, instances of non-cooperation during 6:00-14:00, indicate that IPE profits are limited to the upper boundary $max (u_{m,t})$ (according to {\emph {Lemma 4}}). Note that cooperative strategies include not only financial incentives but also profit limits. Additionally, non-cooperation is significantly less frequent in Case 3. It reveals that the contribution-based MIPDMS is more beneficial to facilitating cooperation. 

\begin{figure}[htb]
	\vspace{-0.2cm}
	\centering
	\includegraphics[scale=0.225]{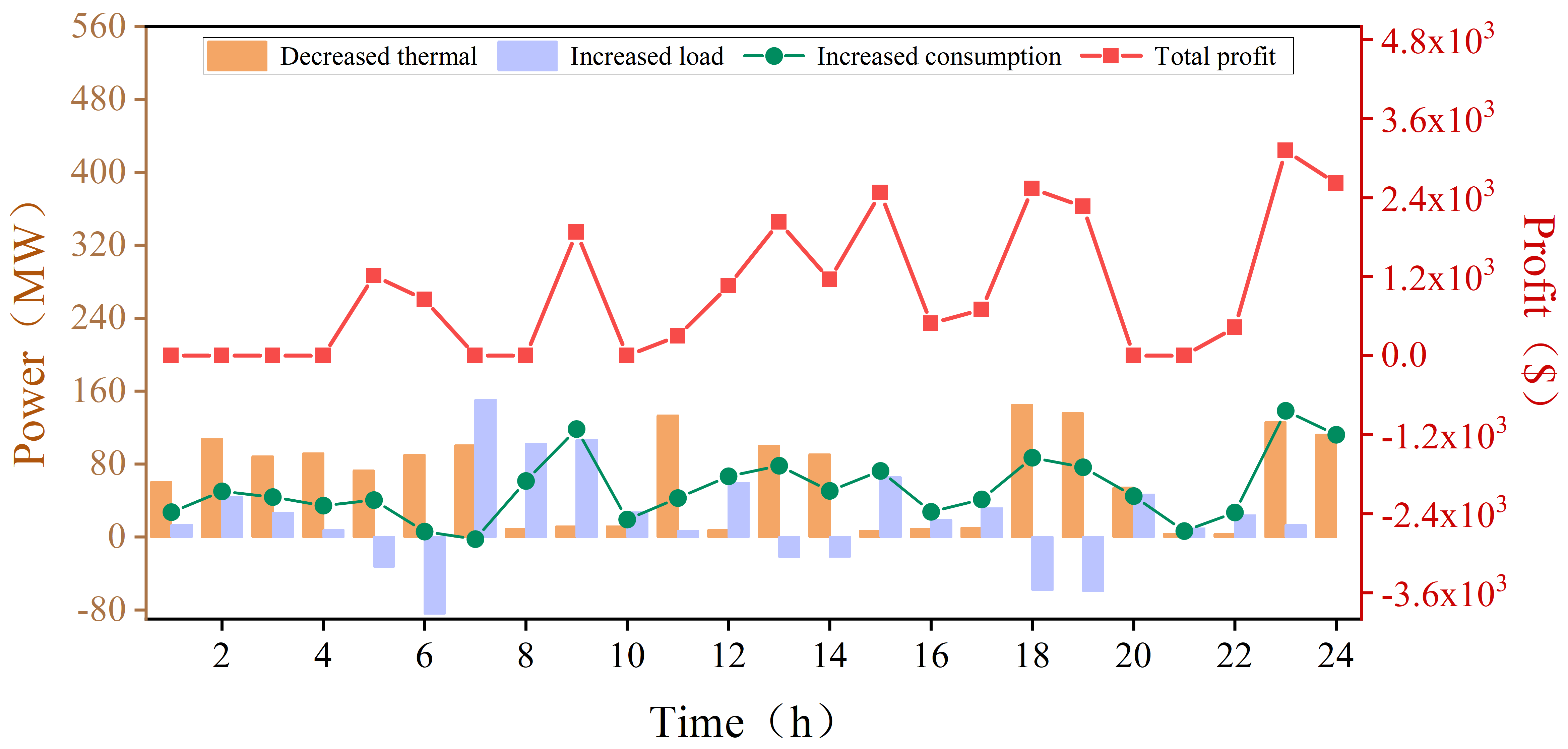}
	\vspace{-0.0cm}
	\caption{Specific response and profit curve in Case 2.}\label{Fig15}
	\vspace{-0.3cm}
\end{figure}

\begin{figure}[htb]
	\vspace{-0.0cm}
	\centering
	\includegraphics[scale=0.225]{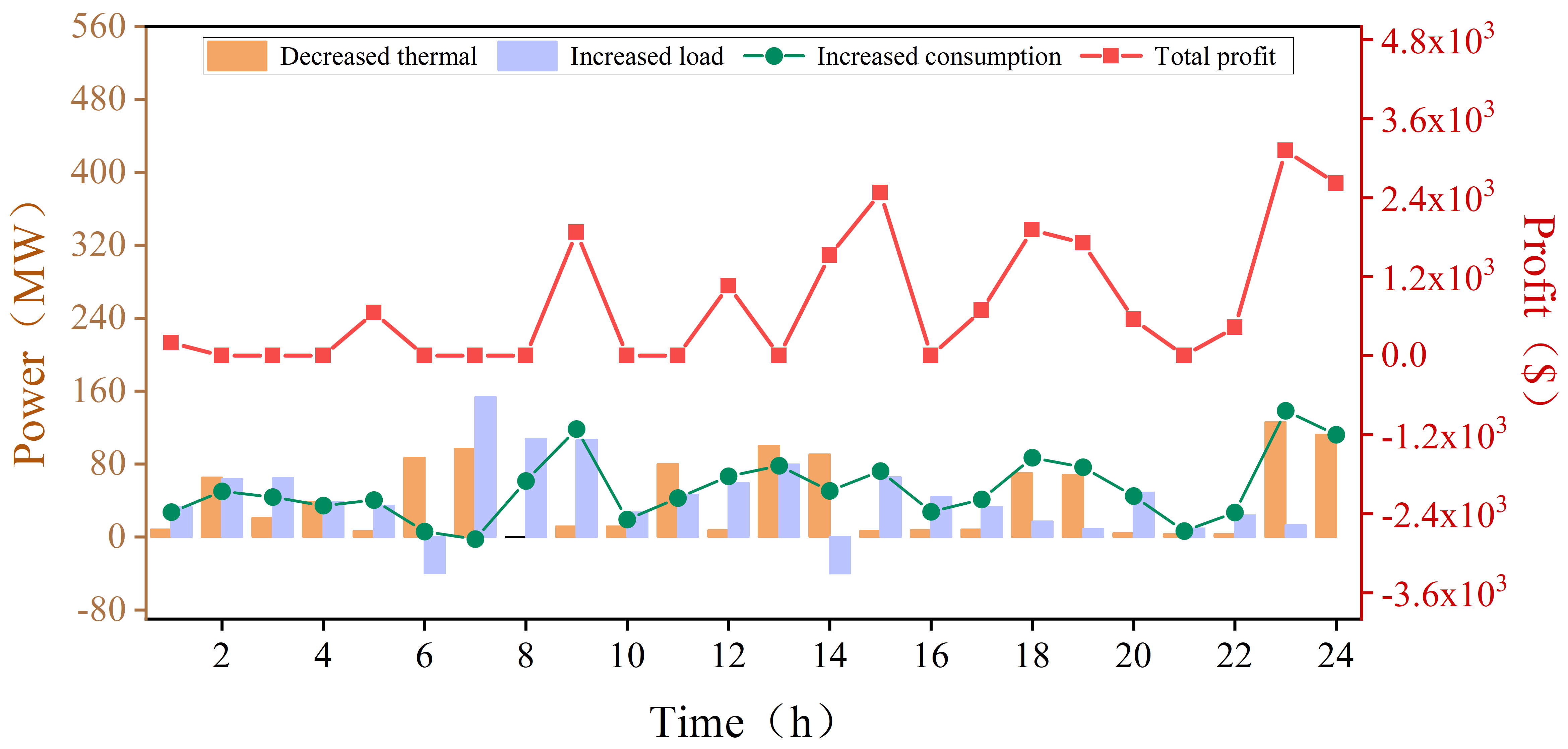}
	\vspace{-0.3cm}
	\caption{Specific response and profit curve in Case 3.}\label{Fig16}
	\vspace{-0.5cm}
\end{figure}

\subsection{Yantai 26 Bus Power System}

The modified Yantai 26-bus power system, a real system in China, is used for further study. In this system, the higher proportion of wind power is connected to buses 8 and 20, and the industrial load connects to bus 22. Furthermore, model parameters of the PMP system also follow the data in Subsection A, and the minimum production quantity is set as 20. The specific network topology and system parameters are provided in Appendix D.

For the day-ahead PDSCR strategy, the Pareto solutions are shown in Fig. \ref{Fig18}, which regards the two upper-level objectives of $\rm C_{ASC}$ (optimal boundary: 0.8053) and total thermal cost (optimal boundary: 81003.65$\$$), as well as wind power curtailment corresponding each Pareto solution. The responding results for the compromise solution are shown in Fig. \ref{Fig19}. Note that in the peak periods of wind power supply, such as 3:00-14:00 in Fig. \ref{Fig19}(a), industrial power demand is in a high level in Fig. \ref{Fig19}(b). Also, the thermal power is correspondingly minimized. Moreover, in the peak periods of wind power supply, the demand curve is close to the wind power supply, which shows that the day-ahead PDSCR is reasonable and satisfactory.

\begin{figure}[!h]
	\vspace{-0.2cm}
	\centering
	\includegraphics[scale=0.223]{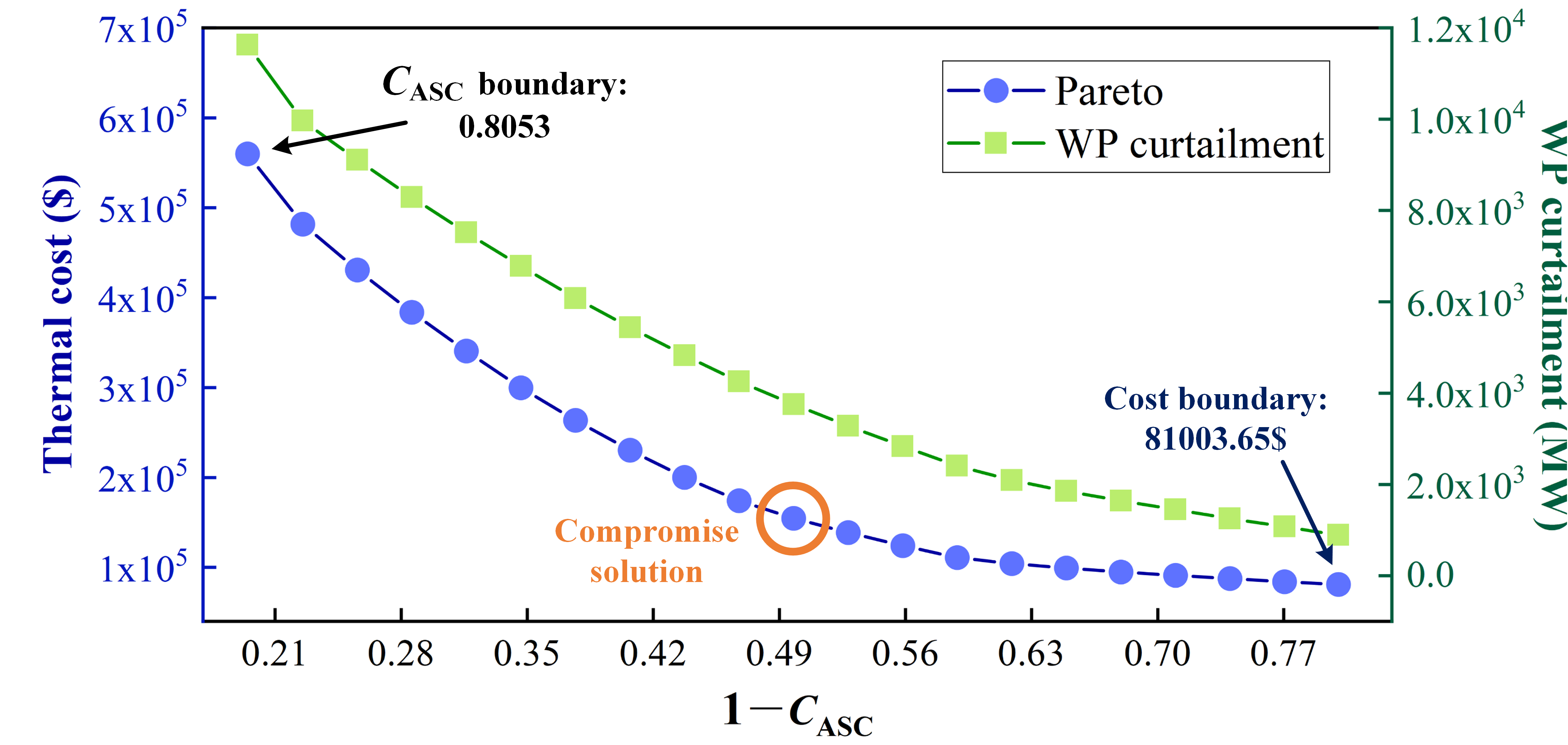}
	\vspace{-0.2cm}
	\caption{PF and corrsponding WP curtailment in Yantai system.}\label{Fig18}
	\vspace{-0.1cm}
\end{figure}

\begin{figure}[!h]
	\vspace{-0.2cm}
	\centering
	\includegraphics[scale=0.263]{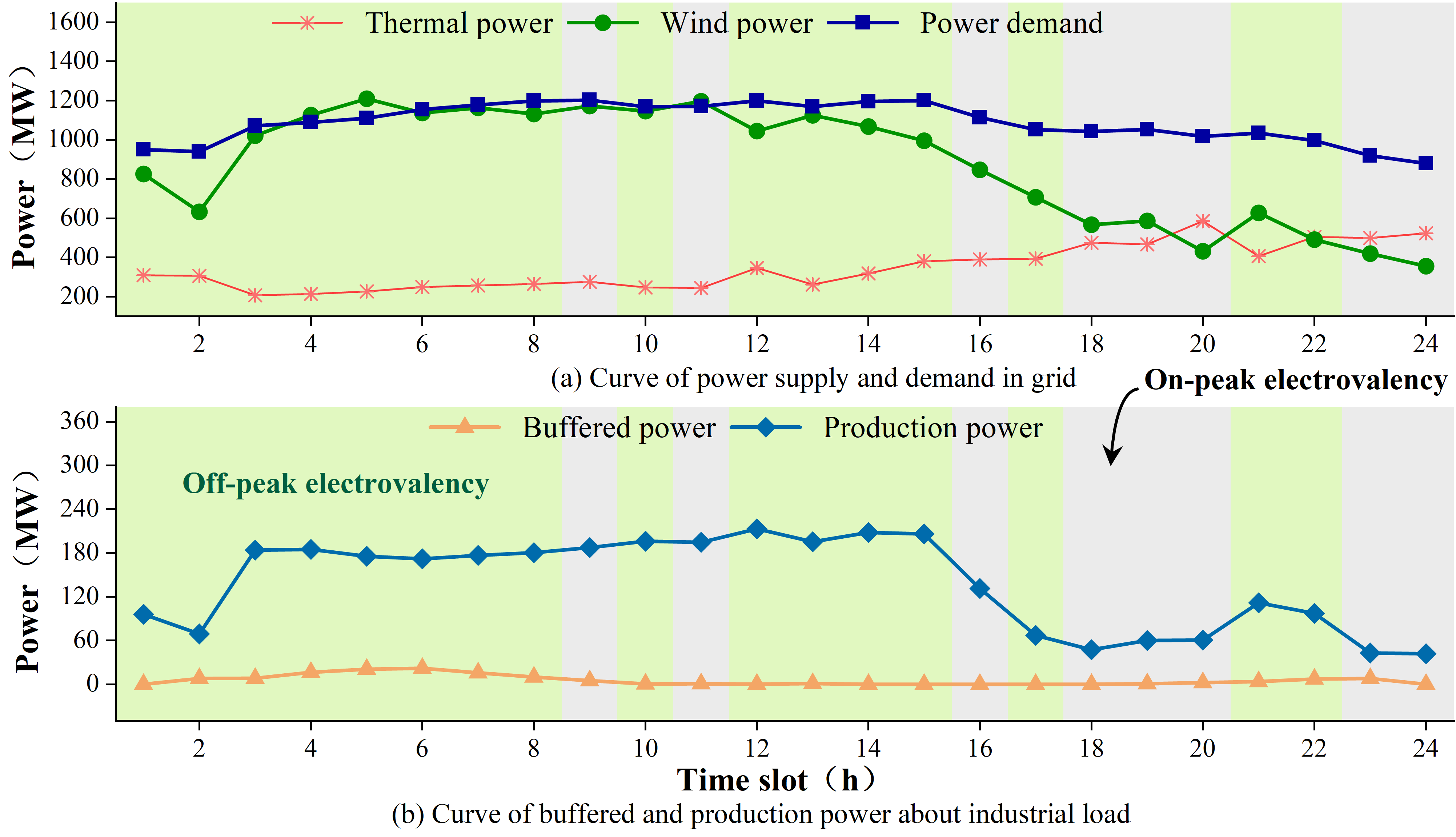}
	\vspace{-0.2cm}
	\caption{Dispatching result in day-ahead.}\label{Fig19}
	\vspace{-0.3cm}
\end{figure}

Moreover, we also simulate the three cases in TABLE III, to further investigate the impact of the equal and contribution on responding performance. Here, a random scenario close to the standard wind power value is selected to make analysis, wind power fluctuation of which is about 301.23 MW. The corresponding simulation results are shown in TABLE V. Similar to the results in TABLE IV, intraday PDSCR for Cases 2 and 3 are conducive to reduce wind power curtailment and demand-supply total cost. It verifies the effectiveness of intraday PDSCR strategy. Furthermore, the total cost on the demand-supply side in Case 3 is lower than that of Case 2, which reveals that contribution-based cooperation would generate more benefits. Furthermore, the contribution-based MIPDMS in Case 3 has a remarkable HV performance, which reveals that the response is more effective, compared with the equal MIPDMS in Case 2. The reason is that superior profit distribution will lead to more effective responding results in profit-driven cooperation. In the numerical perspective, wind power utilization is positive to cooperative profits, and thus the higher wind power proportion will expand the feasible region of profit utility set $({u_{k,t}},v_t)$. Finally, simulation results verify that the regional expansion of the profit utility set $({u_{k,t}},v_t)$ is more advantageous to the contribution-based cooperation, which will lead to the superior response effect.
\vspace{-0.3cm}
\begin{table}[!h]
   \renewcommand{\arraystretch}{1.5}
	\begin{center}
		\caption{Simulation results in Yantai probabilistic sampling scenarios}
		\begin{tabular}{m{0.8cm}<{\centering}m{1.3cm}<{\centering}m{2.0cm}<{\centering}m{1.3cm}<{\centering}m{1.2cm}<{\centering}} \hline\hline	
			Case &  HV /$10^{-2}$ & WP curtailment /$10^3\cdot$MW  & $\rm C_{ASC}/10^{-2}$ & Total cost $10^5\cdot\$$ \\   
			\hline
			Case 1   & —      & 3.61   & 47.22      & 5.42     \\          
			Case 2   & 35.57   & 3.59   & 46.74      & 4.78   \\ 
			Case 3   & 42.06   & 1.86   & 32.24      & 4.48   \\  
			\hline\hline	
		\end{tabular}
	\end{center}
	\vspace{-0.4cm}
\end{table}




\section{CONCLUSION}

This paper proposes day-ahead and intraday PDSCR strategies to promote RE utilization. The corresponding constrained optimization models are established to achieve the cooperative response according to characteristics of the two dispatch periods. Numerical case studies on a modified IEEE 24-bus benchmark and a real power system yield the following conclusions.

({\emph {i}) The proposed PDSCR model and its corresponding strategies are effective in improving wind power utilization. For standard prediction scenarios, the response corresponds exactly to the demand in the day-ahead PDSCR strategy, and intraday PDSCR demonstrates adaptability for uncertain wind scenarios.

({\emph {ii}) In the day-ahead PDSCR, the utility center effectively coordinates thermal units operation and industrial demand shifting, in order to boost the wind power utilization.

({\emph {iii}) In the day-ahead PDSCR, the utility center can well coordinate the operation of thermal units and industrial demand shifting, which also helps enhance wind power utilization.

{ On the other hand, the proposed PDSCR strategy does not consider the integration of energy storage systems. However, there is an increasing need for high-density and large-scale energy storage systems in practice. Our future work will focus on studying and quantifying the impact of energy storage systems on PDSCR. Furthermore, the practical implementation of the response strategy depends on precise data acquisition and robust communication systems \cite{Pref1}. As the demands for data sampling and communication continue to grow, unforeseen events, such as time delays and cyber attacks, are becoming more prevalent. These events can potentially lead to emergency situations in renewable power system. Therefore, the future research will also place more emphasis on considering cyber attacks and their impacts on the our proposed strategy.}





\appendices

\section{Detail PMP constrain model and corrsponding Karush-Kuhn-Tucker conditions}\label{A}

\setcounter{equation}{0}

\renewcommand\theequation{A.\arabic{equation}}

In this Appendix, we will introduce the detail PMP constrain model and corrsponding Karush-Kuhn-Tucker conditions in IPE cost optimization \cite{ref19}. Especially, the basic law is noted as follows: \emph{every project is required to be processed through individual procedures in sequence and it is continuously differentiable}. Therefore, the production behavior of the above PMP system obeys the following principles, where the dual variables of inequalities are $[\lambda _1^{t,i},\lambda _2^{t,i},\lambda _3^{t,i},\lambda _4^{t,i},\lambda _5^{t,i},\lambda _6]$ and the dual variables of equalities are $[{\nu _1^{t,i}},{\nu _2},{\nu _3}]$.

\emph{(i)} Firstly, the number of workpieces processed in the procedure ${\rm P}_i$ during the $t^{th}$ ($t=1, 2,…, {\rm T}$) time slot should satisfy the following boundary constraint.
\begin{equation}\label{eq1}
{{\lambda _1^{t,i}}: {-N_{{\rm{P}},t,i}} \le 0, \quad {\lambda _2^{t,i}}: {N_{{\rm{P}},t,i}} \le N_{{\rm{P}},t,i}^{\max }}
\end{equation}
where ${N_{{\rm{P}},t,i}}$ is the quantity of processed projects of procedure ${{\rm{P}}_i}$ at the end of ${t}$. Furthermore, its maximum value is represented by ${N_{{\rm{P}},t,i}^{\max }}$ . Similarly, the storage quantity of buffer ${{\rm{B}}_i}$, denoted as ${N_{{\rm{B}},t,i}}$, should satisfy the following constraint.
\begin{equation}\label{eq2}
{{\lambda _3^{t,i}}: {-N_{{\rm{B}},t,i}} \le 0, \quad {\lambda _4^{t,i}}: {N_{{\rm{B}},t,i}} \le N_{{\rm{B}},t,i}^{\max }}
\end{equation}
where ${N_{B,i}^{\max }}$ represents the corresponding upper bound.

\emph{(ii)} The projects processed in procedure ${P_i}$ indeed come from the previous buffers or machines, i.e., its number is limited by that of the previous buffer ${{N_{B,t,i}}}$ and previous procedure ${{N_{{\rm{P}},t,i}}}$ for current ${N_{{\rm{P}},t+1,i+1}}$.
\begin{equation}\label{eq3}
{{\lambda _5^{t,i}: {N_{{\rm{P}},t+1,i+1}} - {N_{{\rm{B}},t,i}} - N_{{\rm{P}},t,i}^{\max }} \le 0}
\end{equation}

Also, the total production should achieve the number of workpieces with minimum requirements ${N_{{\rm{TAR}}}}$ as (\ref{eq4}).
\begin{equation}\label{eq4}
{\lambda _6: {N_{{\rm{TAR}}}}-\sum\limits_{t = 1}^{\rm{T}} {{N_{{\rm{P}},t,{\rm{R}}}}}\le 0}
\end{equation}

Furthermore, the workpiece balance of buffer ${{\rm{B}}_i}$ in two successive time slots is expressed as in (\ref{eq5}).
\begin{equation}\label{eq5}
{ {\nu _1^{t,i}}: {N_{{\rm{B}},t,i}} + {N_{{\rm{P}},t ,i}} - {N_{{\rm{P}},t ,i + 1}} - {N_{{\rm{B}},t + 1,i}} = 0}
\end{equation}

\emph{(iii)} The production cycle usually starts/ends without workpieces in machines or buffers, for safety and efficiency concerns. Therefore, this issue is formulated as follows.
\begin{equation}\label{eq6}
{{\nu _2}: \sum\limits_{i = 1}^{{\rm{R}} - 1} {{N_{{\rm{P}},{\rm{T}},i}}}  + \sum\limits_{i = 1}^{{\rm{R}} - 1} {{N_{{\rm{B}},{\rm{T}},i}}}  = 0}
\end{equation}
\begin{equation}\label{eq7}
{{\nu _3}: \sum\limits_{i = 2}^{\rm{R}} {{N_{{\rm{P}},1,i}}}  + \sum\limits_{i = 1}^{{\rm{R}} - 1} {{N_{{\rm{B}},1,i}}}  = 0}
\end{equation}

Based on the constraints (A.1)-(A.7), the convex economic dispatch model of IPE is formulated as follows.
\begin{equation}\label{eq8}
\begin{array}{l}
\begin{aligned}
{\rm{min }} \quad &{J_{{\rm{PMP}}}} = \sum\limits_{t = 1}^{\rm{T}} {\left( {{F_{{\rm{P}},t}} + {F_{{\rm{B}},t}}} \right)}  + {{\rm{C}}_{{\rm{FIXED}}}}\\
{\rm{s.t.}} \quad &{F_{{\rm{P}},t}} = \sum\limits_{i = 1}^{\rm{R}} {{\alpha _t} \cdot {N_{{\rm{P}},t,i}}}  \cdot {c_{{\rm{P}},i}}\\
&{F_{{\rm{B}},t}} = \sum\limits_{i = 1}^{{\rm{R}} - 1} {{\alpha _t} \cdot {N_{{\rm{B}},t,i}}}  \cdot {c_{{\rm{B}},i}} 
\end{aligned}
\end{array}
\end{equation}
where ${c_{P,i}}$ and ${c_{{\rm{B}},i}}$ are cost coefficients of ${P_i}$ and ${B_i}$, ${\alpha _t}$ is electricity price in $t$ slot, and ${{\rm{C}}_{{\rm{FIXED}}}}$ is fixed cost. In the continuous convex model as (\ref{eq8}), it could derive the explicit optimal solution or optimality constraints\cite{ref18}. For the constrained optimization model (A.1)-(A.8), it is continuous and linear when considering parameterized electricity price ${\alpha _t}$. In order to explore explicit optimal solution or optimality constraints, we firstly formulate the corresponding {\emph {Lagrange}} dual function, as shown in (\ref{eq255}), and prove its strong duality by continuous and linear condition.
\begin{equation}\label{eq255}
\begin{array}{l}
\begin{aligned}
{\cal L} =  F(x,y) + {\upsilon ^ \top } \cdot h(x,y) + {\lambda ^ \top } \cdot g(x,y),
\end{aligned}
\end{array}
\end{equation}
where variable $x$ is parameterized, $y$ is the optimization variable, $\upsilon$ and $\lambda$ are dual variables in constraints involving $y$, $h(x,y)$ is the equality constraint, and $g(x,y)$ is the inequality constraint. On this basis, the corresponding sufficient condition for strong duality is presented as follows.

{\noindent{\textbf {Sufficient condition}}: For the constant $\forall  \theta \in [0,1]$ and $\forall x_0$, if partially parameterized ${\cal L}(x,y){|_{x = {x_0}}}$ follows: (continuous and linear condition)}
\begin{equation}\label{eq266}
\begin{array}{l}
\begin{aligned}
{\mathscr L}({x_0},\theta &\cdot{y_1} + (1 - \theta )\cdot{y_2})\\ 
&= \theta  \cdot {\mathscr L}({x_0},{y_1}) + (1 - \theta ) \cdot {\mathscr L}({x_0},{y_2}),
\end{aligned}
\end{array}
\end{equation}
there will be no gap between the \emph {Lagrange} dual optimal ${\cal L}^*$ and the original optimal $F^*$, i.e., ${{\cal L}^*} = {F^*}$ (Slater conditions).

{\noindent \emph{\textbf {Proof}}: Set $A$ is defined as $A = \{ (m,n,k)|\;\exists y \in {\mathop{\rm int}}\; D,\;g({x_0},$ $y) \le m,\;h({x_0},y) = n,F({x_0},y) \le k\}$. In the premise of (\ref{eq26}), $g({x_0},y)$, $h({x_0},y)$ and $F({x_0},y)$ are linear or affine for unique variable $y$ in ${\cal L}(x,y){|_{x = {x_0}}}$, and the corresonding set $A$ is convex. In addition, another convex set $B$ is defined as $B = \{ (0,0,t)|t < {F^*}\}$. It is noted that $A$ and $B$ are separate, because the result ${F^*}({x_0},y) \le t < {F^*}$ caused by intersection $A = B = \{ (0,0,t)\}$ conflicts with the ${F^*}$.

For sets $A$ and $B$, there exists separating hyperplane $(\bar \lambda ,\bar \nu ,\mu ,$ $\gamma ) \ne 0$ to establish (\ref{Ineq1}).
\begin{equation}\label{Ineq1}
\begin{array}{l}
\begin{aligned}
{\bar \lambda ^ \top } \cdot m + {\bar \nu ^ \top } \cdot n + \mu  \cdot k  \ge \gamma ,\; \bar \lambda  \cdot 0 + \bar \nu  \cdot 0 + \mu  \cdot t \le \gamma ,
\end{aligned}
\end{array}
\end{equation}
where $\bar \lambda \succ  0$ and $\mu  \ge 0$, otherwise, the formula (\ref{Ineq1}) will be unboundary. Accordingly, for $\forall t \le {F^*}$, $\mu  \cdot t \le \gamma $ and
\begin{equation}\label{Ineq2}
\begin{array}{l}
\begin{aligned}
\mu  \cdot F + {\bar \upsilon ^ \top } \cdot h + {\bar \lambda ^ \top } \cdot g \ge \gamma  \ge \mu  \cdot {F^*},
\end{aligned}
\end{array}
\end{equation}
i.e., $\mathop {\min }\limits_y {\cal L}{|_{\lambda  = \bar \lambda /\mu ,\nu  = \bar \nu /\mu }} \ge {F^*}$. On the other hand, because of $g \le 0,$ and $\lambda  \ge 0$, $\mathop {\min }\limits_y {\cal L} \le {F^*}$. Therefore, $\mathop {\min }\limits_y {\cal L} = {d^*} = {F^*}$, and there will be no gap between the \emph {Lagrange} dual optimal ${\mathscr L}^*$ and the original optimal $F^*$, i.e., Slater condition established ${{\cal L}^*} = {F^*}$.

$\hfill\blacksquare$

To this end, the specific {\emph {Lagrange}} function about the dispatch of the PMP system is established as (\ref{eq9}), where the weighted sum for dual variables and constraints is regarded as the penalties to primeval object function. In the above convex optimization, there is no duality gap between the primeval problem and its dual, i.e., the sufficient and necessary in KKT conditions is satisfied. Therefore, the optimality constraints of PMP schedules are able to be established based on KKT conditions. The constraints about the gradient of the {\emph {Lagrange}} function, complementary slackness, and the boundary of duel variables are as (\ref{eq10})-(\ref{eq15}).
\begin{equation}\label{eq9}
\begin{array}{l}
\begin{aligned}
{{\cal L}_{{\rm{PMP}}}} &= \sum\limits_{t = 1}^{\rm{T}} {\left( {\sum\limits_{i = 1}^{\rm{R}} {{\alpha _t} \cdot {N_{{\rm{P}},t,i}}}  \cdot {c_{{\rm{P}},i}} + \sum\limits_{i = 1}^{{\rm{R}} - 1} {{\alpha _t} \cdot {N_{{\rm{B}},t,i}}}  \cdot {c_{{\rm{B}},i}}} \right)} \\
 + &\sum\limits_{t = 1}^{\rm{T}} {\sum\limits_{i = 1}^{\rm{R}} {\lambda _1^{t,i} \cdot ({N_{{\rm{P}},t,i}} - N_{{\rm{P}},t,i}^{\max })} }  - \sum\limits_{t = 1}^T {\sum\limits_{i = 1}^{\rm{R}} {\lambda _2^{t,i} \cdot {N_{{\rm{P}},t,i}}} } \\
 + &\sum\limits_{t = 1}^{\rm{T}} {\sum\limits_{i = 1}^{{\rm{R}} - 1} {\lambda _3^{t,i} \cdot ({N_{{\rm{B}},t,i}} - N_{{\rm{B}},t,i}^{\max })} }  - \sum\limits_{t = 1}^{\rm{T}} {\sum\limits_{i = 1}^{{\rm{R}} - 1} {\lambda _4^{t,i} \cdot {N_{{\rm{B}},t,i}}} } \\
 + &\sum\limits_{t = 1}^{{\rm{T}} - 1} {\sum\limits_{i = 1}^{{\rm{R}} - 1} {\lambda _5^{t,i} \cdot ({N_{{\rm{P}},t + 1,i + 1}} - {N_{{\rm{B}},t,i}} - N_{{\rm{P}},t,i}^{\max })} } \\
 + &{\lambda _6} \cdot ({N_{{\rm{TAR}}}} - \sum\limits_{t = 1}^{\rm{T}} {{N_{{\rm{P}},t,{\rm{R}}}}} ) + {{\rm{C}}_{{\rm{FIXED}}}}\\
 + &\sum\limits_{t = 1}^{{\rm{T}} - 1} {\sum\limits_{i = 1}^{\rm{R}} {\nu _1^{t,i} \cdot ({N_{{\rm{B}},t,i}} + {N_{{\rm{P}},t,i}} - {N_{{\rm{P}},t,i + 1}} - {N_{{\rm{B}},t + 1,i}})} } \\
 + &{\nu _2} \cdot (\sum\limits_{i = 1}^{{\rm{R}} - 1} {{N_{{\rm{P}},{\rm{T}},i}}}  + \sum\limits_{i = 1}^{{\rm{R}} - 1} {{N_{{\rm{B}},{\rm{T}},i}}} )\\
 + &{\nu _3} \cdot (\sum\limits_{i = 2}^{\rm{R}} {{N_{{\rm{P}},1,i}} + \sum\limits_{i = 1}^{{\rm{R}} - 1} {{N_{{\rm{B}},1,i}}} )} 
\end{aligned}
\end{array}
\end{equation}
\begin{equation}\label{eq10}
{{\nabla _{{N_{{\rm{P}},t,i}}}}{{\cal L}_{{\rm{PMP}}}} = 0,\quad {\nabla _{{N_{{\rm{B}},t,i}}}}{{\cal L}_{{\rm{PMP}}}} = 0}
\end{equation}
\begin{equation}\label{eq11}
\lambda _1^{t,i} \cdot ({N_{{\rm{P}},t,i}} - N_{{\rm{P}},t,i}^{\max }) = 0, \quad \lambda _2^{t,i} \cdot {N_{{\rm{P}},t,i}} = 0
\end{equation}
\begin{equation}\label{eq12}
\lambda _3^{t,i} \cdot ({N_{{\rm{B}},t,i}} - N_{{\rm{B}},t,i}^{\max }) = 0, \quad \lambda _4^{t,i} \cdot {N_{{\rm{B}},t,i}} = 0
\end{equation}
\begin{equation}\label{eq13}
\lambda _5^{t,i} \cdot ({N_{{\rm{P}},t,i}} - {N_{{\rm{B}},t - 1,i - 1}} - N_{{\rm{P}},t - 1,i - 1}^{\max }) = 0
\end{equation}
\begin{equation}\label{eq14}
{\lambda _6} \cdot ( - \sum\limits_{t = 1}^T {{N_{{\rm{P}},t,{\rm{R}}}} + {N_{{\rm{TAR}}}}} ) = 0
\end{equation}
\begin{equation}\label{eq15}
\lambda _1^{t,i},\lambda _2^{t,i},\lambda _3^{t,i},\lambda _4^{t,i},\lambda _5^{t,i},{\lambda _6} \ge 0
\end{equation}
where the variables, electricity price ${\alpha _t}$, included in the decision of utility center are uncontrollable for IPE. It makes the optimal solution space of PMP economic dispatch parameterized.

\section{Supplemental Constraints of Day-ahead and Intraday PDSCR strategy}\label{B}

\setcounter{equation}{0}

\renewcommand\theequation{B.\arabic{equation}}

This Appendix introduces the decision variables and supplemental constraints in day-ahead and intraday PDSCR strategy, and it is divided into two-step.

{\emph{\textbf{Step 1:}}}
\begin{equation}\label{Apeq7}
\begin{array}{l}
\begin{aligned}
&x=[{s_{t,p}},{P_{{\rm T},t,p}},{R_{{\rm T},t,p}},{P_{{\rm W},t,p}},{{\alpha _t}}]\\[1mm]
&y=[{N_{{\rm P},t,i}},{N_{{\rm{B}},t,i}}]
\end{aligned}
\end{array}
\end{equation}

The decision variables $x$ about grid include thermal on/off status ${s_{t,p}}$, power outputs ${P_{{\rm T},t,p}}$, reserves ${R_{{\rm T},t,p}}$, wind power utilization ${P_{{\rm W},t,l}}$ and electricity price ${{\alpha _t}}$ for time $t$, thermal generator $p$, wind unit $l$, number of thermal power units $N_{\rm G}$. $y$ about IPE includes processed projects ${N_{{\rm P},t,i}}$ and buffer ${N_{{\rm{B}},t,i}}$ for time $t$ and process $i$. $F({P_{{\rm T},t,i}})$ and $S_{{\rm T},t,p}$ present generation cost function and starting cost as (\ref{Apeq8}) and (\ref{Apeq9}), respectively. Here, $a$, $b$, $c$ denote the fuel cost coefficients of thermal unit. It is noted that $S_{{\rm T},t,p}$ is constant for each thermal generator, which is quantified by the given start-up cost coefficients ${\alpha _p}$, ${\beta _p}$, ${\tau _i}$ and down time $T_{t,i}^{{\rm{off}}}$ \cite{refre7}. On this basis, the corresponding constraints are as follows.
\begin{equation}\label{Apeq8}
F({P_{{\rm T},t,p}}) = (aP_{{\rm T},t,p}^2 + b{P_{{\rm T},t,p}} + c) \cdot {s_{t,p}}
\end{equation}
\begin{equation}\label{Apeq9}
{S_{{\rm{T}},t,p}} = \left[ {{\alpha _p} + {\beta _p}\left( {1 - {e^{\frac{{ - T_{t,p}^{{\rm{off}}}}}{{{\tau _p}}}}}} \right)} \right] \cdot {s_{t,p}}
\end{equation}
where $F({P_{{\rm T},t,p}})$ is quadratic for $P_{{\rm T},t,p}$ and introduced the 0-1 variables of thermal on/off status ${s_{t,p}}$ as (\ref{Apeq10}), which make the grid dispatch problem non-convex.
\begin{equation}\label{Apeq10}
{s_{t,p}} = \left\{ {\begin{array}{*{20}{c}}
{1,\quad {''}{\rm {on}}}{''}\\
{0,\quad {''}{\rm {off}}}{''}
\end{array}} \right.
\end{equation}

For On/Off status determined by start and stop action, the duration about each start to stop and stop to strate should be separately longer than the minimum operating and rested duration ${M_{{\rm{up}},p}}$ and ${M_{{\rm{dn}},p}}$ as (\ref{Apeq11}), where ${S_{up,p}}$ and ${S_{{\rm dn},p}}$ are continuous time periods as (\ref{Apeq12}).
\begin{equation}\label{Apeq11}
{S_{up/down,p}} \ge {M_{{\rm{up/down}},p}}
\end{equation}
\begin{equation}\label{Apeq12}
{S_{up/down,p,t}} = \left\{ {\underbrace {0, \cdots ,0}_{t - 1},\underbrace {1, \cdots ,1}_{{S_{{\rm{up/down}},p}}},\underbrace {0, \cdots ,0}_{T - {S_{{\rm{up/down}},p}} - t + 1}} \right\}
\end{equation}

The thermal power limits and climb constraints are formulated as (\ref{Apeq15}) and (\ref{Apeq16}), i.e., power boundaries and the changing rate.
\begin{equation}\label{Apeq15}
{P_{{\rm{T}},p,min}} \le {P_{{\rm{T}},t,p}} \le {P_{{\rm{T}},p,{\rm{max}}}}
\end{equation}
\begin{equation}\label{Apeq16}
 - {r_{d,p}}\Delta t \le {P_{{\rm{T,}}t,p}} - {P_{{\rm T},t - 1,p}} \le {r_{u,p}}\Delta t
\end{equation}

For wind power, its connection should be less than available supply as (\ref{Apeq17}) in each wind turbine, and the total wind power ${P_{{\rm{f}},t}}$ could be denoted as (\ref{Apeq18}).
\begin{equation}\label{Apeq17}
0 \le {P_{{\rm W},t,l}} \le P_{{\rm W},t,l}^{{\rm{predict}}}
\end{equation}
\begin{equation}\label{Apeq18}
{P_{{\rm{f}},t}} = \sum\limits_{i = 1}^{{N_{\rm{W}}}} {{P_{{\rm W},t,l}}} 
\end{equation}

For thermal power ${P_{T,t,p}}$, wind power ${P_{W,t,l}}$ , industial load and traditional load ${D_{Tra,t}}$, the adjustment should follow the power balance as (\ref{Apeq19}).
\begin{equation}\label{Apeq19}
\begin{array}{l}
\sum\limits_{i = 1}^{{N_{\rm G}}} {{P_{{\rm T},t,p}}}  + \sum\limits_{p = 1}^{{N_{\rm{W}}}} {{P_{{\rm W},t,l}}}  = {D_{{\rm Tra},t}}  \\
+ \sum\limits_{i = 1}^{\rm R} {{N_{{\rm{P}},t,i}}}  \cdot {c_{{\rm{P}},i}} + \sum\limits_{i = 1}^{{\rm R} - 1} {{N_{{\rm{B}},T,i}}}  \cdot {c_{{\rm{B}},i}}
\end{array}
\end{equation}

The reserve should accommodate the error of wind power and load as (\ref{Apeq20}), where $\beta$ persents the error coefficient.
\begin{equation}\label{Apeq20}
\begin{array}{l}
\begin{aligned}
&\beta \sum\limits_{i = 1}^{{N_{\rm{W}}}} {P_{W,t,l}^{{\rm{Fore}}}} \le \sum\limits_{l = 1}^{{N_{\rm T}}} {{R_{{\rm T},t,p}}}
\le \sum\limits_{p = 1}^{{N_{\rm{G}}}} {{P_{{\rm{T}},i,max}}} + \sum\limits_{i = 1}^{{N_{\rm{W}}}} {{P_{{\rm W},t,l}}}\\
&  - {D_{{\rm Tra},t}} - \sum\limits_{i = 1}^{\rm R} {{N_{{\rm{P}},t,i}}}  \cdot {c_{{\rm{P}},i}} + \sum\limits_{i = 1}^{\rm R - 1} {{N_{{\rm{B}},T,i}}}  \cdot {c_{{\rm{B}},j}}
\end{aligned}
\end{array}
\end{equation}

Furthermore, constraints on DC power flow are shown in (\ref{Apeq21}), where ${P_{{\rm{G}}i}}$ and ${P_{Di}}$ are the generation and demand power, ${\theta _i}$ and ${N_i}$ denote phase angle of the $i^{th}$ bus and the number of connected buses. Moreover, ${B_{ij}}$ is the transfer matrix between bus $i$ and $j$. Besides, ${s_{\rm{f}}}$ and ${s_{\rm{P}}}$ represent numbers of wind farms and PMP systems. The risk of power transmission is considered via the limit of transmitted power as in (\ref{Apeq22}) \cite{rereref4}.
\begin{equation}\label{Apeq21}
{P_{{\rm{G}}i}} = {P_{Di}} - {s_{\rm{f}}}{P_{\rm{f}}} + {s_{\rm{P}}}\sum\limits_{t = 1}^{\rm{T}} {E{C_t}}  + \sum\limits_{j \in {N_i}} {{B_{ij}}({\theta _i} - {\theta _j})}
\end{equation}
\begin{equation}\label{Apeq22}
\left| {T{P_{t,j}}} \right| \le T{P_{j,\max }} \cdot (1 - {C_{\rm ASC}})
\end{equation}

{\emph{\textbf{Step 2:}}}
\begin{equation}\label{Apeq23}
\begin{array}{l}
\begin{aligned}
&{{x} = [P_{{\rm T},t,g}^{{\rm{UP}}},P_{W,t,l}^{{\rm{UP}}},P_{outres}^t]} \\[1mm]
&{{y}=[{N_{{\rm P},t,i}},{N_{{\rm{B}},t,i}}]}
\end{aligned}
\end{array}
\end{equation}

The decision variables $x_{des2}$ about grid include thermal power output ${P_{{\rm T},t,g}^{{\rm{UP}}}}$, wind power utiliztion ${P_{{\rm W},t,l}}$ and outside reserve power ${{P_{\rm outres}^t}}$ for time $t$ and thermal generator $g$, and wind power unit $l$. ${y_{des2}}$ about IPE includes processed projects ${N_{{\rm P},t}^{\rm{UP}}}$ and buffer ${N_{{\rm{B}},t}^{\rm{UP}}}$ for time $t$ and process $i$. The corresponding constraints are as follows.
\begin{equation}\label{Apeq24}
C_{\rm outres}^t = \lambda P_{\rm outres}^t
\end{equation}
where $C_{outres}^t$ denotes purchasing cost of external power, and $\lambda$ denotes the corresponding purchasing price.
\begin{equation}\label{Apeq27}
P_{{\rm{WP,cur}}}^t = P_{{\rm W},t}^{{\rm{actual}}} - \sum\limits_{l = 1}^{{N_{\rm{W}}}} {P_{{\rm W},t,l}^{{\rm{UP}}}}
\end{equation}
where $P_{{\rm{WP,cur}}}^t$ denotes the curtailment of wind power in slot $t$, $P_{{\rm W},t}^{{\rm{actual}}}$ denotes actual total wind power in slot $t$ and $P_{{\rm W},t,l}^{{\rm{UP}}}$ denotes actual wind power connection in wind unit $i$ and slot $t$. For thermal power ${P_{{\rm{T}},t,g}^{\rm UP}}$, wind power ${P_{{\rm W},t,l}^{UP}}$ , reserve power $P_{{\rm outres}}^t$, industrial load and traditional load ${D_{{\rm Tra},t}}$, the adjustment should follow the power balance as (\ref{Apeq28}).
\begin{equation}\label{Apeq28}
\begin{array}{l}
\sum\limits_{g = 1}^{{N_{\rm G}}} {P_{{\rm{T}},t,g}^{\rm UP}}  + \sum\limits_{l = 1}^{{N_{\rm{W}}}} {P_{{\rm W},t,l}^{\rm UP}}  + P_{{\rm outres}}^t =  \\
{D_{{\rm Tra},t}} + \sum\limits_{i = 1}^{\rm R} {{N_{{\rm{P}},t,i}^{\rm {UP}}}}  \cdot {c_{{\rm{P}},i}} + \sum\limits_{i = 1}^{R - 1} {{N_{{\rm{B}},T,i}^{\rm {UP}}}}  \cdot {c_{{\rm{B}},i}}
\end{array}
\end{equation}

The adjustment constraints considering thermal power reserves are as (\ref{Apeq29}).
\begin{equation}\label{Apeq29}
{P_{{\rm{T}},t,g}} - {R_{{\rm{T}},t,g}} \le P_{{\rm{T}},t,g}^{\rm {UP}} \le {P_{{\rm{T}},t,g}} + {R_{{\rm{T}},t,g}}
\end{equation}
where the information ${P_{T,t,g}}$ and ${R_{T,t,g}}$ are based on step 1 scheme. 


\section{Proof of Convexity In Profit Utility Function }\label{C}

\setcounter{equation}{0}

\renewcommand\theequation{C.\arabic{equation}}

{\emph{\textbf{Lemma 1:}}} For $(u,v)$, $u\in {S_1} : \{ u = {x^ \top }Ax + {c_1}^ \top x|a_1^ \top x \le {b_1}, a_2^ \top x = {b_2},u \in {\mathbb{R}^n}\}, v\in {S_2} : \{ v = {c_2}^ \top y|a_3^ \top y \le {b_3},a_3^ \top y = {b_3},y \in \mathbb{R}{^m}\} $, set $S : \{ (u,v)|{u_i},{v_j} > 0,u \in \mathbb{R}{^n},v \in \mathbb{R}{^m}\} $ is convexity, where $A = diag\{ {a_1},{a_2},...,{a_n}\} $. 

{\emph{\textbf{Proof:}}} For classic convex set as halfspace ${C_1} : \{ x|a_1^ \top x $ $\le {b_1},x \in \mathbb{R}{^n}\} $ and hyperplane ${C_2} : \{ x|a_2^ \top x = {b_2},x \in \mathbb{R}{^n}\} $, $\{ x|{C_1} \cap {C_2},x \in \mathbb{R}{^n}\}$ is convex. Similarly, for ${C_3} = \{ y|a_3^ \top y \le {b_3},y \in \mathbb{R}{^m}\}$ and ${C_4} = \{ y|a_4^ \top y \le {b_4},y \in \mathbb{R}{^m}\}$, intersection $\{ y|{C_3} \cap {C_4},y \in \mathbb{R}{^m}\}$ is convex. 

Obviously, the linear mapping $u_i^{'}$ for $x$ defined as $u_i^{'} = {c_1}^ \top x$, is continuous and convex in $\{ x|{C_1} \cap {C_2},x \in \mathbb{R}{^n}\}$. Further, the quadratic mapping $u_i^{''}$ for $x$ defined as $u_i^{''} = {x^ \top }Ax$ , the conditions are as (\ref{Apeq1}) according to convexity definition.
\begin{equation}\label{Apeq1}
\left\{ \begin{array}{l}
 {\theta _1} \cdot x_1^ \top A{x_1} + (1 - {\theta _1}) \cdot x_2^ \top A{x_2} = x_3^ \top A{x_3}\\
 {\theta _2} \cdot {x_1} + (1 - {\theta _2}) \cdot {x_2} = {x_3}\\
 0 \le {\theta _1},{\theta _2} \le 1
\end{array} \right.
\end{equation}
where $x_1,x_2 \in \{ x|{C_1} \cap {C_2},x \in \mathbb{R}{^n}\} $. When matrix $A$ is diagonalized as $diag\{a_{1},a_{2},...,a_{n}\}$, for each $x_{1,i}, x_{2,i}, x_{3,i}$, the above equation will be equivalent to (\ref{Apeq2}).
\begin{equation}\label{Apeq2}
\left\{ \begin{array}{l}
 {\theta _1} \cdot x_{1,i}^2 + (1 - {\theta _1}) \cdot x_{2,i}^2 = x_{3,i}^2\\
 {\theta _2} \cdot {x_{1,i}} + (1 - {\theta _2}) \cdot {x_{2,i}} = {x_{3,i}}\\
 0 \le {\theta _1},{\theta _2} \le 1
\end{array} \right.
\end{equation}
where the quadratic could be reduced as follows.
\begin{equation}\label{Apeq3}
({\theta _1} - \theta _2^2) \cdot x_{1,i}^2 + 2(\theta _2^2 - {\theta _2}){x_{1,i}}\cdot{x_{2,i}} - ({\theta _1} + \theta _2^2 - 2{\theta _2}) \cdot x_{2,i}^2 = 0
\end{equation}

In (\ref{Apeq3}), the quadratic discriminant is established, i.e. for $({\theta _1},{\theta _2})\in\mathbb{R}, 4{({\theta _1} - {\theta _2})^2} \ge 0$. Therefore, $u_i^{''}$ is also continuous and convex in $\{ x|{C_1} \cap {C_2},x \in \mathbb{R}{^n}\}$. Further, it is noted that the convex direct product $S_1^{'} \times {S_2}$ and $S_1^{''} \times {S_2}$ preserve convexity, where $S_1^{'}$ and $S_1^{''}$ are defined as $u_i^{'}$ and $u_i^{''}$ separately.

Then, we discuss the convex about partial sum $S = \{ (u,v)|u = {u^{'}} + {u^{''}},({u^{'}},v) \in S_1^{'},({u^{''}},v) \in S_1^{''}\},S_1^{'},S_1^{''} \in \mathbb{R}{^n} \times \mathbb{R}{^m}$. For $0 \le \theta  \le 1$, the conditions are as (\ref{Apeq4}) according to convexity definition.
\begin{equation}\label{Apeq4}
\begin{array}{l}
\theta  \cdot ({u_1},{v_1}) + (1 - \theta )({u_2},{v_2})\\
 = (\theta  \cdot (u_1^{'} + u_1^{''}) + (1 - \theta ) \cdot (u_2^{'} + u_2^{''}), \theta  \cdot {v_1} + (1 - \theta ) \cdot {v_2})
\end{array}
\end{equation}

Furthermore, 
\begin{equation}\label{Apeq5}
(\theta  \cdot u_1^{'} + (1 - \theta ) \cdot u_2^{'}, \theta  \cdot {v_1} + (1 - \theta ) \cdot {v_2}) \in S_1^{'}
\end{equation}
\begin{equation}\label{Apeq6}
(\theta  \cdot u_1^{''} + (1 - \theta ) \cdot u_2^{''}, \theta  \cdot {v_1} + (1 - \theta ) \cdot {v_2}) \in S_1^{''}
\end{equation}

Therefore, $\theta  \cdot ({u_1},{v_1}) + (1 - \theta )({u_2},{v_2}) \in S$, which is equivalent to $S = \{ (u,v)|{u_i},{v_j} > 0,u \in \mathbb{R}{^n},v \in \mathbb{R}{^m}\}$, is convex.

$\hfill\blacksquare$

\section{Yantai 26-buses System}\label{D}

{As shown in Fig. \ref{Fig17}, the network topology depicts the connectivity status of all buses. Detailed network parameters can be found in the GitHub files:} \url{https://github.com/Xinxin-Long/YANTAI_SYSTEM_FILE.git}.
\begin{figure}[!h]
	\vspace{-0.2cm}
	\centering
	\includegraphics[scale=0.78]{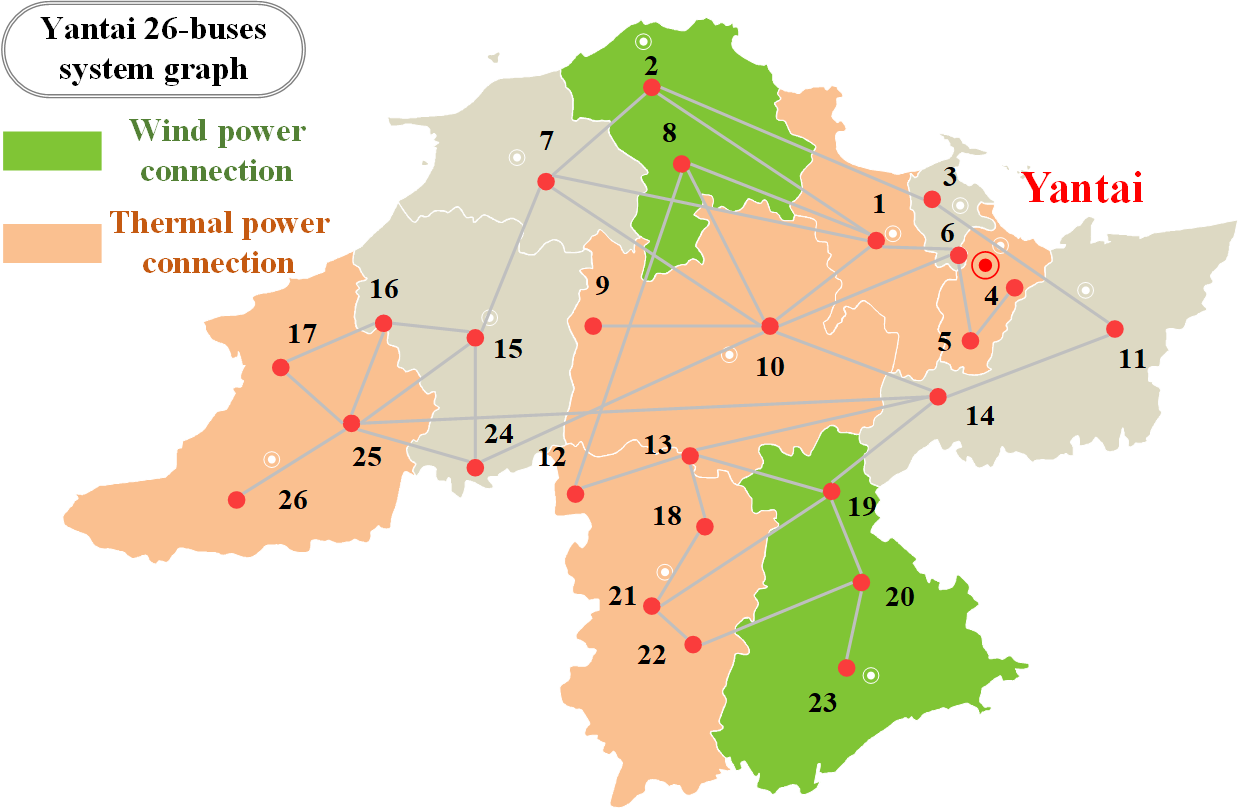}
	\vspace{0cm}
	\caption{Yantai 26 bus power system.}\label{Fig17}
	\vspace{-0.3cm}
\end{figure}

\bibliographystyle{IEEEtran}
\bibliography{PDSCR.bbl}

\begin{IEEEbiography}[{\includegraphics[width=1in,height=1.25in,clip,keepaspectratio]{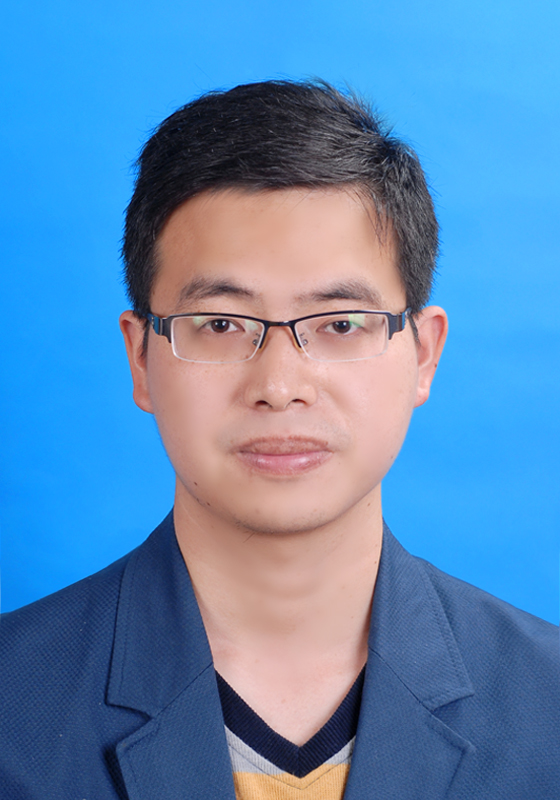}}]{Yuanzheng Li} received the M.S. degree and Ph.D. degree in Electrical Engineering from Huazhong University of Science and Technology (HUST), Wuhan, China, and South China University of Technology (SCUT), Guangzhou, China, in 2011 and 2015, respectively. Dr Li is currently an Associate Professor in HUST. He has published several peer-reviewed papers in international journals. His current research interests include electric vehicle, deep learning, reinforcement learning, smart grid, optimal power system/microgrid scheduling and decision making, stochastic optimization considering large-scale integration of renewable energy into the power system and multi-objective optimization.
\end{IEEEbiography}


\begin{IEEEbiography}[{\includegraphics[width=1in,height=1.25in,clip,keepaspectratio]{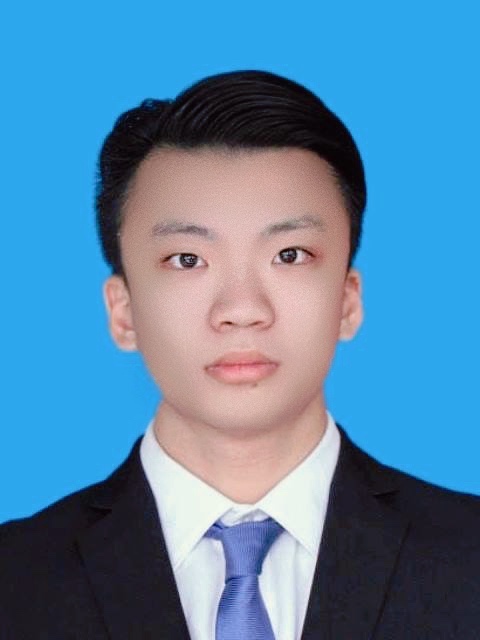}}]{Xinxin Long} received the B.S. degree in Electrical Engineering from the Guangxi University, Nanning, China, in 2021. He is currently working toward the M.S. degree in Huazhong University of Science and Technology (HUST), Wuhan, China. His research interests include optimal power system scheduling, decision making with uncertainties, convex optimization, and other application of artificial intelligence in Smart Grid.
\end{IEEEbiography}


\begin{IEEEbiography}[{\includegraphics[width=1in,height=1.25in,clip,keepaspectratio]{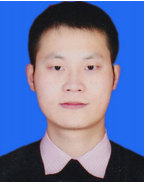}}]{Yang Li} received the Ph.D. degree in electrical engineering from North China Electric Power University, Beijing, China, in 2014. He is an Associate Professor with the School of Electrical Engineering, Northeast Electric Power University, Jilin, China. He is currently also a China Scholarship Council-funded Postdoc with Argonne National Laboratory, Lemont, IL, USA. His research interests include power system stability and control, integrated energy systems, renewable energy integration, and smart grids.
\end{IEEEbiography}


\begin{IEEEbiography}[{\includegraphics[width=1in,height=1.25in,clip,keepaspectratio]{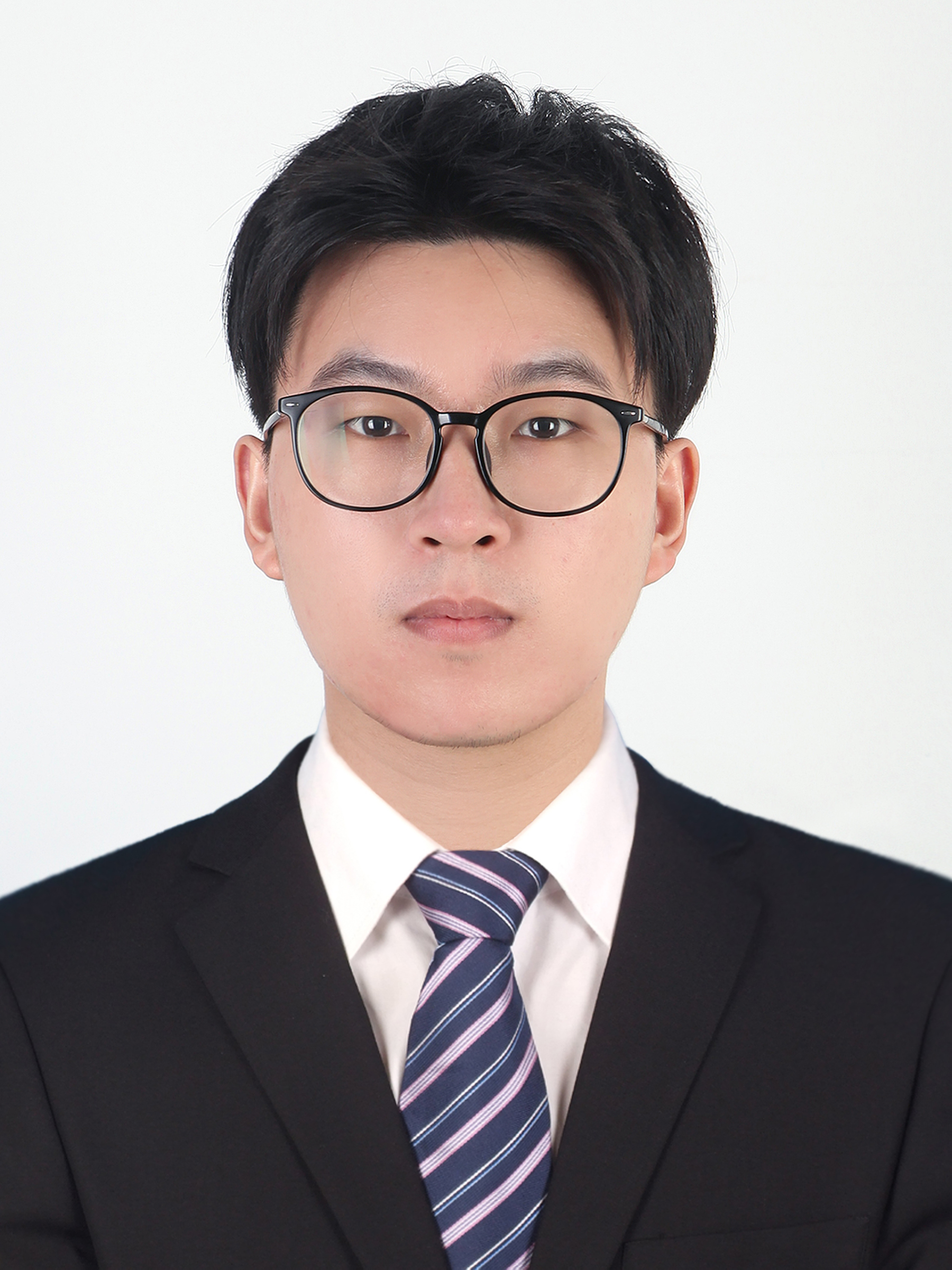}}] {Yizhou Ding} received the B.S. degree in communication
engineering from the Wuhan University of Technology (WHUT) in 2021. He is currently working toward the M.S. degree in Huazhong University
of Science and Technology (HUST), Wuhan,
China. His research interests include electricity price forecast, renewable power system operation, deep learning, federated learning, and graph neural network, and other application of artificial intelligence in Smart Grid.
\end{IEEEbiography}

\begin{IEEEbiography}[{\includegraphics[width=1in,height=1.25in,clip,keepaspectratio]{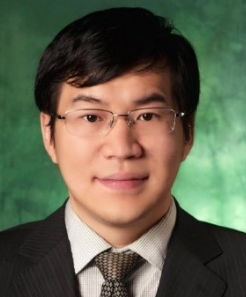}}]{Tao Yang} received the Ph.D. degree in electrical engineering from Washington State
University, Pullman, WA, USA, in 2012.
He is currently an Assistant Professor with
the Department of Electrical Engineering, University of North Texas (UNT), Denton, TX, USA.
Between August 2012 and August 2014, he was
an ACCESS Postdoctoral Researcher with the
ACCESS Linnaeus Centre, Royal Institute of
Technology, Sweden. Prior to joining the UNT in
2016, he was a Scientist/Engineer with Energy
$\&$ Environmental Directorate, Pacific Northwest National Laboratory. His research interests include distributed control and optimization with applications to power systems, cyber-physical systems, networked control systems, and multiagent systems.
Dr. Yang received the Ralph E. Powe Junior Faculty Enhancement Award in 2019 from Oak Ridge Associated Universities and the Best Student Paper Award (as an advisor) at the 14th IEEE International Conference on Control and Automation.
\end{IEEEbiography}


\begin{IEEEbiography}[{\includegraphics[width=1in,height=1.25in,clip,keepaspectratio]{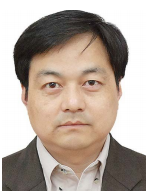}}]{Zhigang Zeng} (IEEE Fellow) received the Ph.D. degree in systems analysis and integration from Huazhong University of Science and Technology,Wuhan, China, in 2003.
			He is currently a Professor with the School of Automation and the Key Laboratory of Image Processing and Intelligent Control of the Education Ministry of China, Huazhong University of Science and Technology. He has published more than 100 international journal articles. His current research interests include theory of functional differential equations and differential equations with discontinuous right-hand sides and their applications to dynamics of neural networks, memristive systems, and control systems. Dr. Zeng was an Associate Editor of the IEEE TRANSACTIONS ON NEURAL NETWORKS AND LEARNING SYSTEMS from 2010 to 2011. He has been an Associate Editor of the IEEE TRANSACTIONS ON CYBERNETICS since 2014 and the IEEE TRANSACTIONS ON FUZZY SYSTEMS since 2016, and a member of the Editorial Board of Neural Networks since 2012, Cognitive Computation since 2010, and Applied Soft Computing since 2013.
\end{IEEEbiography}
\end{document}